\documentclass[12pt,preprint]{aastex}

\def\hi   {\protect\ion{H}{1}}

\slugcomment{Accepted for publication by The Astronomical Journal}

\shorttitle{NGC 1311 Clusters}
\shortauthors{Eskridge et al.}

\begin{document}

\title{Star Clusters in the Nearby Late-Type Galaxy NGC 1311\footnotemark[1]}

\author{Paul B.~Eskridge\altaffilmark{2}}
\email{paul.eskridge@mnsu.edu}

\author{Richard de Grijs\altaffilmark{3,4}, Peter Anders\altaffilmark{5}, 
Rogier A.~Windhorst\altaffilmark{6}, Violet A.~Mager\altaffilmark{7,8}, \& 
Rolf A.~Jansen\altaffilmark{6}} 

\footnotetext[1]{Based on observations with the NASA/ESA Hubble Space 
Telescope, obtained at the Space Telescope Science Institute, which is operated 
by the Association of Universities for Research in Astronomy, Inc.~under NASA 
contract No.~NAS5-26555.}
\altaffiltext{2}{Department of Physics and Astronomy, Minnesota State
University, Mankato, MN 56001}
\altaffiltext{3}{Department of Physics and Astronomy, University of Sheffield, 
Sheffield S3 7RH, United Kingdom}
\altaffiltext{4}{National Astronomical Observatories, Chinese Academy of 
Sciences, Beijing, P.R.~China}
\altaffiltext{5}{Sterrenkundig Instituut, Universiteit Utrecht, NL-3508 TA 
Utrecht, The Netherlands}
\altaffiltext{6}{School of Earth \& Space Exploration, Arizona State 
University, Tempe, AZ 85287}
\altaffiltext{7}{Department of Physics and Astronomy, Arizona State University,
Tempe, AZ 85287}
\altaffiltext{8}{Observatories of the Carnegie Institution of Washington, 
Pasadena, CA 91101}

\begin{abstract}
Ultraviolet, optical and near infrared images of the nearby ($D \approx 5.5$ 
Mpc) SBm galaxy NGC 1311, obtained with the {\it Hubble Space Telescope}, 
reveal a small population of 13 candidate star clusters.  We identify candidate 
star clusters based on a combination of their luminosity, extent and spectral
energy distribution.  The masses of the cluster candidates range from 
$\sim$10$^3$ up to $\sim$10$^5$ $M_{\odot}$, and show a strong positive trend
of larger mass with increasing with cluster age.  Such a trend follows from the 
fading and dissolution of old, low-mass clusters, and the lack of any young 
super star-clusters of the sort often formed in strong starbursts.  The cluster 
age distribution is consistent with a bursting mode of cluster formation, with 
active episodes of age $\sim$10 Myr, $\sim$100 Myr and $\ga$1 Gyr.  The ranges 
of age and mass we probe are consistent with those of the star clusters found 
in quiescent Local Group dwarf galaxies.
\end{abstract}

\keywords{galaxies: individual (NGC 1311) --- galaxies: spiral --- galaxies: 
star clusters --- infrared: galaxies --- ultraviolet: galaxies}

\section{Introduction}

Star clusters are powerful tools for probing the star-formation history and 
chemical enrichment of galaxies (e.g., \citeauthor{h61}\citeyear{h61};
\citeauthor{whit99}\citeyear{whit99}; \citeauthor{dk2}\citeyear{dk2}; 
\citeauthor{grijs}\citeyear{grijs}).  In nearby systems, star clusters provide 
us with spatially resolved examples of simple stellar populations (SSPs).  
Observations of individual cluster stars are, therefore, crucial for 
understanding the effect of metallicity on stellar evolution (e.g., 
\citeauthor{carme}\citeyear{carme} and references therein).  In more distant 
systems, star clusters provide us with examples of unresolved SSPs.  Analysis
of statistical samples of such clusters offers us a wealth of information on 
the history of star formation in their parent galaxies (e.g., 
\citeauthor{grijs}\citeyear{grijs} and references therein).  Understanding star 
and cluster formation in a range of environments is essential for understanding 
galaxy evolution because star cluster formation traces the strongest episodes 
of galactic star formation (e.g., \citeauthor{dGNA}\citeyear{dGNA}).  

There has been a great deal of work on star cluster populations and formation 
histories in nearby luminous spiral galaxies over the last decade (e.g., 
\citeauthor{whit99}\citeyear{whit99}; \citeauthor{grijs}\citeyear{grijs}; 
\citeauthor{dk2}\citeyear{dk2}; \citeauthor{lar04}\citeyear{lar04}).  By 
comparison, the star cluster formation history in nearby, low-luminosity 
galaxies has received relatively little attention (but see, e.g., 
\citeauthor{bhe}\citeyear{bhe}; \citeauthor{a04}\citeyear{a04}; 
\citeauthor{seth04}\citeyear{seth04}; \citeauthor{dGA06}\citeyear{dGA06}).  
Studies of nearby galaxies are important because we can resolve abundant 
small-scale detail in them.  Moreover, the low metallicities of low-luminosity 
nearby galaxies offer us a close-up view of the star-formation process that may
better resemble that in the high-redshift, early Universe.

Studies of detailed stellar formation histories are still restricted to fairly 
nearby galaxies, and are most powerful when they combine high angular 
resolution with broad wavelength coverage.  The number of very nearby galaxies 
with such data available is still quite small.  NGC 1311 is a very nearby ($D 
\approx 5.5$ Mpc), but little-studied late-type (SBm) galaxy.  
\citeauthor{tully}(\citeyear{tully}) identify it as a member of the 14$+$14 
Association, a loose group dominated by the luminous spiral NGC 1313.  Table 1 
summarizes the basic properties of the system.  NGC 1311 was a target in 
several {\it Hubble Space Telescope} (HST) snapshot surveys (GO programs 9124 
and 9824; \citeauthor{wind02}\citeyear{wind02}, \citeauthor{vio}\citeyear{vio}, 
and \S 2 below).  As a result, a set of broad-band images spanning a wide
wavelength interval (0.3---1.6$\mu$m) at sub-arcsecond resolution now exists 
for this galaxy.  As NGC 1311 is quite nearby, the bright star clusters and 
luminous individual stars are detected as discrete sources.  We can thus probe 
the spatially resolved star-formation history of NGC 1311 by studying the 
broad-band spectral energy distributions of the star clusters, the individual 
stars, and the unresolved light.  This paper is concerned with the star 
clusters of NGC 1311.  We shall address the individual stars and unresolved 
light in future publications.

In \S 2 we describe the HST observational data.  We present the observed 
properties of the candidate star clusters in \S 3, and our analysis of these 
observations in \S 4.  We summarize our conclusions, and discuss issues for 
further research in \S 5.

\section{Observational Data}

The data for this study are a set of UV, optical and NIR images obtained with 
the Wide-Field and Planetary Camera 2 (WFPC2) and the Near Infrared Camera and
MultiObject Spectrograph (NICMOS) on board HST.  We have WFPC2 images taken 
through the F300W, F606W and F814W filters, and a NICMOS image taken with the
NIC3 camera through the F160W filter.  Details of the observations are given in 
Table 2.  The F300W and F814W WFPC2 images were obtained as part of the HST 
program GO-9124 ``Mid-UV Snapshot Survey of Nearby Irregulars:  Galaxy 
Structure and Evolution Benchmark'' (R.~Windhorst PI).  For these images, the 
nuclei of the target galaxies were placed on the WF3 chip.  Details of the 
observing and reduction procedures for these data are given in 
\citeauthor{wind02}(\citeyear{wind02}).  The F160W image was obtained as part 
of the HST program GO-9824 ``NIC3 SNAPs of Nearby Galaxies Imaged in the 
mid-UV:  The Remarkable Cool Stellar Population in Late-Type Galaxies'' 
(R.~Windhorst PI).  Details of the observing and reduction procedures for these 
data are given in \citeauthor{vio}(\citeyear{vio}).  The archival WFPC2 F606W 
image of NGC 1311 was obtained as part of the HST program GO-9162 ``Local 
Galaxy Flows and the Local Mass Density'' (R.~Tully PI).  The WFPC2 Wide-Field 
Camera (WFC) spatial sampling is $\approx 0{''}\llap.10$ per pixel.  Our search
for candidate star clusters turned up no examples on the WFPC2 Planetary Camera 
CCD, so we do not use the PC data in this study.  The NIC3 spatial sampling is 
$0{''}\llap.20$ per pixel.  We show images of the central 42$'' \times 26''$ of 
NGC 1311 in the four observed bands in Figure 1.

\section{Candidate Star Clusters}

Candidate star clusters in external galaxies can be identified in several ways.
Physically large clusters in sufficiently nearby galaxies are extended sources.
Compact star clusters, or star clusters in more distant galaxies can be 
identified by their luminosity in the most extreme cases.  Such clusters can 
also be distinguished from bright stars by their spectral energy distribution 
(SED).  \citeauthor{whit99}(\citeyear{whit99}), 
\citeauthor{a04}(\citeyear{a04}, \citeyear{a07}) and 
\citeauthor{bas05}(\citeyear{bas05}) give more detailed discussions of how to 
distinguish star clusters from individual luminous stars.  We make use of both 
luminosity and SED criteria to define our list of candidate star clusters in 
NGC 1311. 

We used HSTPhot\footnotemark[7] (\citeauthor{hstpht}\citeyear{hstpht}) to
extract stellar photometry from the WFPC2 images  For the NICMOS image, we used
the version of DAOPHOT (\citeauthor{hat87}\citeyear{hat87}) embedded in the 
XVISTA image analysis package (\citeauthor{stov88}\citeyear{stov88}).  We then 
selected candidate clusters by searching the output photometry files for 
sources that are either extended compared to the point-spread function (PSF), 
or have colors that deviate from those of individual stars.  We discuss the 
details of our analysis below.

\footnotetext[7]{We used the May 2003 revision of HSTPhot v.~1.1.5b obtained 
from {\tt http://purcell.as.arizona.edu/wfpc2\_calib/}.}

\subsection{Stellar Photometry}

HSTPhot is designed to perform stellar photometry on WFPC2 images, including 
aperture corrections, charge-transfer efficiency corrections, and zero-point 
calibrations.  The zero-points to the VEGAMAG system are thus updated from 
those of \citeauthor{holtz}(\citeyear{holtz}).  For the NICMOS image, we first 
applied the non-linearity correction determined by 
\citeauthor{nonlin}(\citeyear{nonlin}).  This refines the zero-point 
calibration from the 2004 June standard \citep{nic70}.  We then extracted 
stellar photometry with XVISTA/DAOPHOT.  For the NICMOS photometry, we had to 
determine the aperture correction manually by measuring the asymptotic count 
rates for two bright, isolated stars in the observed field.  The correction 
from the 2-pixel radius to infinite aperture is 0.10$\pm$0.02 magnitudes.  All 
of our photometry is calibrated to the VEGAMAG system.  

We compare our stellar photometry to isochrones derived from the models of 
\citeauthor{thiso}(\citeyear{thiso}).  In order to plot the isochrones with the 
photometry, we need an estimate of the distance to NGC 1311.  The best
available distance estimate for NGC 1311 is that of 
\citeauthor{tully}(\citeyear{tully}), who quote $D = 5.45$ Mpc based on the
magnitude of the tip of the red giant branch.  There appear to be no 
metallicity measurements of NGC 1311 in the literature.  For the distance 
quoted above, we find that the isochrones with $Z=0.004$ provide the best 
visual fit to the CMDs.  The more metal-poor isochrones have upper main 
sequences that are substantially bluer than observed in all available colors.  
The isochrones more metal-rich than $Z=0.008$ do not predict stars luminous 
enough in the ultraviolet compared to the data.  We defer a more detailed 
discussion of the isochrone fitting to our paper on the stellar photometry.  We 
adopt a metallicity of $Z=0.004$ below unless otherwise noted.

\subsection{Identification of Cluster Candidates}

There are a number of objects that are not well-fit by the isochrones in 
color-color diagrams.  Figure 2 shows two color-color diagrams that demonstrate 
the problem.  There are objects that are bluer than the isochrones for 
short-wavelength filter combinations and redder than the isochrones for 
long-wavelength filter combinations.  Several of the oddly colored objects are 
within 3$\sigma$ of the photometric limits in at least one bandpass.  In those 
cases, we adopt the conservative view that the odd colors are likely caused by 
relatively large errors near the photometric limit.  We note that a large 
fraction of the points shown in Fig.~2 deviate from the isochrones.  This is
due to a selection bias.  There are a relatively small number of normal stars
that are bright enough at both 3000\AA\ and at 8000\AA\ to be detected in both
F300W and F814W.  

In Figure 3, we show an example of a stellar color--magnitude diagram (CMD) 
with the objects with secure measurements of their anomalous colors marked with 
large circles and those with uncertain measurements, too near the relevant 
photometric limits, enclosed within large diamonds.  In practice, we impose a 
magnitude limit of $I_{814} < 22$ on objects selected by their photometric 
peculiarity.  This is a simple cut-off to impose, and we found that objects
fainter than this limit have color uncertainties large enough that their
deviation from the isochrones in color-color plots are not statistically
significant.  A number of objects are substantially brighter than this faint 
limit, and still have colors that deviate from the isochrones by more than 
2$\sigma$.  On this basis, we consider them to be candidate star clusters.  
Figure 4 shows a portion of the F606W image with the photometrically selected 
candidate clusters circled.  We note that the faintest cluster candidates so 
identified have apparent magnitudes of $V_{606} \approx 22.6$, or absolute 
magnitudes of $M_{606} \approx -6.1$.  

A stellar cluster with an age of $\sim$10 can have extremely red supergiants 
dominating the red and NIR bands, as well as hot blue main sequence stars 
dominating the blue and NUV bands.  At ages of $\sim$100 Myr, the red spectrum 
can be dominated by luminous asymptotic giant branch stars and the blue 
spectrum by blue supergiants (\citeauthor{p83}\citeyear{p83}; 
\citeauthor{m05}\citeyear{m05}).  This will cause the integrated colors of such 
a cluster to appear blue in bands dominated by the luminous blue stars, and red 
in bands dominated by the luminous red stars; exactly the phenomenon 
demonstrated in Fig.~2, and one that has been well known to apply to Magellanic
Cloud clusters for many years (\citeauthor{p83}\citeyear{p83}).  We note that 
the candidate clusters do not stand out in the CMD (see Fig.~3).  It is only in 
the color-color diagrams that their peculiarity becomes evident.  This is in 
keeping with the results of \citeauthor{bas05}(\citeyear{bas05}), who showed 
that broad wavelength sampling of the SED could discriminate between star 
clusters and luminous single stars.

The HSTPhot output includes a number of parameters that are useful in
distinguishing extended sources from point sources.  These are discussed in
detail by \citeauthor{hstpht}(\citeyear{hstpht}).  The $\chi$ parameter is a 
goodness-of-fit parameter derived from the standard statistical $\chi^2$.  A 
good, well-fit stellar image in an uncrowded field should have $\chi \la 1.5$.  
Sources with $\chi$ much greater than this are candidate star clusters, but can 
also include stellar blends and background galaxies.  The sharpness parameter 
is zero for a perfectly-fit stellar image, negative for objects more extended 
than the PSF, and positive for artifacts that are too sharp (cosmic rays, for 
instance).  A stellar PSF image in a relatively uncrowded field should have a 
sharpness of $0 \pm 0.5$.  A plot of the sharpness parameter, output from 
HSTPhot, against $V_{606}$ is shown in Figure 5.  The figure indicate that 
there are plausible cluster candidates up to a magnitude fainter than the 
candidates found above.  Fainter than $V_{606} \approx 24$ mag, the 
HSTPhot-identified ``extended sources'' are consistent with the outer envelope 
of point sources in the $V_{606}$--sharpness plot.  They are thus not 
significantly extended in our data.  However, there is a group of nine sources 
with $22.4 \leq V_{606} \leq 23.6$ that are clearly offset from the stellar 
locus in this magnitude range (see Fig.~5).  None of these sources have colors 
that distinguish them from the stellar isochrones at a statistically 
significant level.  We indicate these extended sources by diamonds in Figure 6. 
Two of them are obvious background spirals, and two more are likely background 
objects.  These four are labelled ``bg'' in Fig.~6, and shown by large, dark 
crosses in Fig.~5.  The remaining five sources are additional faint candidate 
clusters in NGC 1311, bringing the total to 13.  Postage stamp images from the 
F606W data, centered on the candidate clusters, and oriented as in Fig.~6, are 
shown in Figure 7.  Each postage stamp measures $6{''}\llap{.}5$ square.

We make no formal assessment of our completeness level, as this is not critical
for our current purposes.  However, we point out that we are unlikely to have 
missed any luminous clusters ($V_{606} \la 22$), even in the most crowded part 
of the field.  At fainter magnitudes, our census is certainly incomplete.  Thus
we make no attempt to quantify the cluster formation rate.  We shall address
the completeness function (for point sources) in our paper on the individual
stars in NGC 1311.

\section{Discussion}

\subsection{Extendedness Tests}

We see from Fig.~7 that several of the cluster candidates are quite compact.
For an object at a distance of 5.45 Mpc a WFC pixel subtends only about 2.6 pc. 
Thus unresolved sources are either very compact, or they are 
foreground/background point sources.  The simplest way to test for extendedness 
is to plot the radial profiles of the candidate clusters along with those of 
stars.  We show such plots in Figure 8.  Cluster candidates 1, 5, 6, 10 and 12 
are not significantly more extended than stars in their vicinity.  The 
remaining sources are clearly extended at WFPC2 resolution.  This is consistent 
with the HSTPhot results shown in Fig.~5.  

Globular star clusters in the Milky Way have an average half-light radius
of $R_{hl} \approx 3.3 \pm 1.7$ pc (determined from the database of 
\citeauthor{hcat}\citeyear{hcat}\footnotemark[8], not including the distant 
halo clusters).  At the distance of NGC 1311, this corresponds to
$0{''}\llap{.}12$, or about 1.2 WFC pixels.  Figure 9 shows a histogram of the
observed half-light radii of our candidate clusters, along with those of the 
comparison stars used in Fig.~8.  There is a group of objects that are
marginally resolved (comparable to globular clusters at the distance of NGC
1311), and another group that is much more extended.  The most extended objects
are also the fainter ones (candidates 8,11 and 13).  It may be that these
objects are background galaxies, or statistical fluctuations in the disk of NGC 
1311.  However, they have physical sizes ($R_{hl} \approx 10$pc) and absolute 
magnitudes ($M_V \approx -5$ to $-$7) similar to those of Galactic open 
clusters (see Table 6.2 of \citeauthor{binmer}\citeyear{binmer}).

\footnotetext[8]{We used the February 2003 database revision available at 
\break {\tt http://www.physics.mcmaster.ca/Fac\_Harris/Databases.html}.}

\subsection{Background and Foreground Contamination}

We note that all of the compact candidates were identified as such because 
their colors are unusual for luminous hot stars.  Possible sources of 
contamination are foreground (Galactic) stars, background QSOs, and blending of
stars in NGC 1311.  The five compact sources are all within the 48$'' \times 
80''$ region shown in Fig.~4, and all have $V_{606} \la 22.6$.  The expected 
foreground Galactic stellar contamination down to this magnitude level is less
than one star (\citeauthor{rbm}\citeyear{rbm}).  It seems unlikely from this 
that all five sources are foreground stars.  We shall return to this point when 
we consider photometric modelling of the source colors below.

It is unlikely the sample is contaminated by background QSOs.  Based on the
bright QSO number counts of \citeauthor{kk82} (\citeyear{kk82}), we expect an 
average of 0.07 QSOs with $V_{606} \la 22.6$ in our 48$'' \times 80''$ field
of view.

\subsection{Analysis of Cluster Photometry}

\subsubsection{Aperture Correction}

As noted in \S 3.1 above, we used HSTPhot to perform aperture photometry on our 
WFPC2 data, and measured our own set of aperture corrections for the NICMOS 
photometry.  In both cases we expect excellent photometric results (statistical
errors of $\la 0.05$ mag for sources brighter than 22nd mag) for stellar 
sources.  But even marginally resolved sources, such as compact clusters, will 
require different aperture corrections than individual stars.  While HSTPhot is 
able to deliver excellent photometry for marginally resolved clusters 
(\citeauthor{dk1}\citeyear{dk1}; \citeyear{dk2}), some of our sources are 
clearly well-resolved.  We thus adopted the more general methodology for 
aperture correction of HST/WFPC2 data from 
\citeauthor{apcor}(\citeyear{apcor}).

We began by fitting circular 2-D Gaussians to each of the 13 candidate clusters 
in the F606W image.  We then chose source and sky apertures and cluster light 
profiles for each object.  Typically the source aperture was 1 to 2 pixels 
larger than the measured full width at half-maximum (FWHM) of the Gaussian fit, 
and the sky apertures were 2 (inner) and 4 (outer) pixels larger than the 
source aperture.  Note that this will subtract counts due to both the sky and 
to diffuse light in NGC 1311 that is smooth on angular scales of $\la$4 WFC 
pixels ($0{''}\llap{.}4$).  This will also subtract the extended cluster light, 
however that flux should be accounted for the by aperture correction, as 
discussed below.  

We used the formalism of \citeauthor{apcor}(\citeyear{apcor}) to determine 
aperture corrections for each source.  As \citeauthor{apcor}(\citeyear{apcor}) 
model aperture corrections for the F555W, but not the F606W filter, we must 
assume that there is no significant aperture-correction color term between the 
F555W and F606W filters.  As the characteristic wavelengths of the two filters
differ by only $\approx$650\AA, this is not unreasonable.  For the compact, red 
objects (candidates 1, 7 and 10) we used the King30 profile (a 
\citeauthor{king62}\citeyear{king62} profile with $r_t / r_c = 30$, 
corresponding to a concentration index of 1.48).  This profile is appropriate 
for star clusters that are old enough to be tidally truncated 
(\citeauthor{king62}\citeyear{king62}).  For the bluer objects (all the others) 
we used the EFF15 profile (an \citeauthor{eff87}\citeyear{eff87} profile with a 
power-law index of 1.5), as star clusters in the nearby galaxies are better-fit 
with power-law profiles than with King models (e.g., 
\citeauthor{mg03}\citeyear{mg03}; \citeauthor{dGA06}\citeyear{dGA06} and 
references therein).  

The measured FWHMs from the F300W and F814W frames are statistically consistent 
with those found from the F606W frame, but are less well determined due to the 
lower signal-to-noise ratio in these shallower images.  Given this, we decided 
to apply the F606W aperture corrections to the much shallower F814W and F300W 
data.  This is reasonable unless there is significant mass segregation in the 
target clusters.  For young clusters, mass segregation can cause both the
F300W and F814W profiles to be more concentrated than the F606W profile, due
to the massive main-sequence stars and red supergiants, respectively.  For old
clusters, mass segregation can also cause both the F300W and F814W profiles to 
be more concentrated than the F606W profile, due to the hot horizontal branch 
stars and upper RGB and AGB stars respectively.  In both cases, the F606W 
aperture correction will be an overcorrection for the other bands.  It is
impossible to predict how much of a problem this will be for a given cluster
because of statistical sampling of the stellar mass function (e.g., 
\citeauthor{j01}\citeyear{j01}).  Except for very populous clusters, the number 
of stars in the various evolutionary states in question is very small.  We 
discuss this possibility further in \S 4.3.2, below.

Our NICMOS data were obtained with the NIC3 camera.  There are no 
\citeauthor{apcor}(\citeyear{apcor}) aperture corrections for the NIC3 camera.
However, the large pixel size and PSF of the NIC3 camera makes aperture 
correction a less severe problem than for our WFPC2 data.  Our 2-pixel radius 
aperture has an angular radius of $\approx 0{''}\llap{.}4$, or about 4 WFC 
pixels.  This is larger than the F606W FWHM for most of the cluster candidates. 
The exceptions are candidates 2, 8, 11 and 13.  Candidate 8 is undetected in 
the NICMOS image.  For the other candidates, we measured the flux at radii 
larger than 2 NIC3 pixels.  The inclusion of this flux would amount to an 
additional correction that is small compared to the standard aperture 
correction (0.1 mag).  We thus make no further adjustments to the aperture 
corrections for our NIC3 data.

In Table 3 we present our photometric measurements for each of the candidate 
star clusters.  Column 1 is a serial ID.  Column 2 shows the unprocessed (X,Y) 
pixel locations on individual chips of the F606W frame in the upper row, and 
the J2000.0 Right Ascension and declination in the lower row.  Column 3 shows 
which chip the cluster is located on in the F606W frame.  The upper rows of 
columns 4--7 give the apparent magnitudes in F300W (\emph{UV}$_{300}$), F606W 
($V_{606}$), F814W ($I_{814}$) and F160W ($H_{160}$) for the cluster 
candidates, and columns 8--13 give the resulting colors.  The errors on these 
quantities are given in the lower rows.  

\subsubsection{Modelling of Cluster Photometry}

We have analysed the photometry for the candidate clusters that are detected in
all four passbands with AnalySED (\citeauthor{clus03}\citeyear{clus03}; 
\citeauthor{clus04}\citeyear{clus04}).  For our adopted distance and 
metallicities of $Z=0.004$ and 0.008, this results in a set of ages, 
extinctions, and masses for these candidate star clusters.  As noted in \S 3.1,
above, the $Z=0.004$ are generally a better match to the single-star 
photometry, and metallicities larger than 0.008 are clearly inconsistent with
the observed F300W magnitudes.  We built models with both metallicities to
bracket the plausible values for the star clusters.  The results are given in 
Table 4.  We generated fits using both the \citeauthor{ccm}(\citeyear{ccm}) and 
the \citeauthor{calz}(\citeyear{calz}) extinction laws.  The results for the 
two extinction laws are generally consistent with one another at the 1$\sigma$ 
level.  The values and errors quoted in Table 4 reflect the range of results 
for the two extinction laws.  

The ages from AnalySED fall into three broad ranges.  The youngest clusters 
have ages of $\la$10 Myr (\#s 3 \& 5).  There is an intermediate age group with 
ages of $\sim$100 Myr (\#s 1, 4, 6 \& 12).  Finally, there are two clusters 
with ages $\ga$1 Gyr (\#s 7 \& 10).  We note that the photometry for cluster 7 
is consistent with a much younger age if $Z=0.008$.  However, such a young age 
also requires an extinction that is both much higher than the foreground 
extinction (\citeauthor{sfd98}\citeyear{sfd98}), and the typical internal
extinction for low-luminosity late-type galaxies.  Furthermore, the images
of NGC 1311 (see Fig.~1) show no evidence for regions of strong localized 
extinction.  For these reasons, we believe the age result for the $Z=0.004$ is
more plausible for cluster 7.  

In Figure 10, we plot cluster age against cluster mass from the AnalySED 
results for $Z=0.004$ (the plot for $Z=0.008$ is similar).  We include the
candidates detected in only two passbands (see Table 3).  There is a clear
positive trend.  Both the Spearman rank and the Kendall's $\tau$ tests return
probabilities of correlation of $>$99\%, despite the small sample.  The line 
is a simple linear bisector, showing that the data are well-described by a
power law.  We hasten to point out that several effects can contribute to the
appearence of such a plot, and that the resulting fit parameters are, 
therefore, unlikely to have a single physical interpretation:  The scarcity of 
old, low-mass clusters may reflect the detection limit of the sample, but can 
also follow from the dynamics of cluster dissolution (e.g., 
\citeauthor{gbl}\citeyear{gbl}).  NGC 1311 shows no signs of the massive, young 
super star-clusters that form in strong starbursts (e.g., 
\citeauthor{whit99}\citeyear{whit99}).  Its recent cluster formation history is 
similar to that of other undisturbed, low-luminosity, late-type galaxies (e.g., 
\citeauthor{bhe}\citeyear{bhe}; \citeauthor{m07}\citeyear{m07}).

We can also study the properties of the candidate clusters, including those
that are only detected in a subset of our pass-bands, with less data-intensive
techniques.  In Figure 11 we present two color-color diagrams of the cluster 
candidates, along with predictions of the \citeauthor{thiso}(\citeyear{thiso}) 
SSP models.  The models shown in Fig.~11 all have Solar heavy-element abundance 
ratios, and include convective overshooting.  Comparisons with 
$\alpha$-enhanced models do not change the qualitative results shown in 
Fig.~11.  Most of the candidate clusters are consistent with low-metallicity 
models with ages in the range 3--100 Myr.  Two cluster candidates (\#s 4 \& 12) 
have colors that lie off all the SSP models.  Consideration of all the 
color-color plots indicate that both these objects are either too red compared 
to SSP models (that is, they are too bright in both F814W and F160W), or too
bright in F300W.  This argues against a calibration systematic, as the problem 
appears in both the WFPC2 and NICMOS data.  Both of these objects were 
initially flagged as candidate clusters based on their colors.  One of the two 
objects (\#12) is unresolved, and the other (\#4) is marginally resolved.  This 
suggests that, perhaps, these are cases of close blends of bright stars with 
significantly different colors.  It is also possible that these objects are
examples of clusters that have experienced significant mass-segregation.  As
noted in \S 4.3.1, above, we expect that the F606W aperture corrections would
{\it overcorrect} the F300W magnitudes for mass-segregated clusters.  This
could lead to the blueward displacement that we find for these clusters in
Fig.~11.

Two cluster candidates (\#s 7 \& 10) have implied ages of $\ga$1 Gyr from both 
the AnalySED fits and the \citeauthor{thiso}(\citeyear{thiso}) SSP models.  
The observed colors are consistent with \citeauthor{thiso}(\citeyear{thiso}) 
model ages as large as those of ancient Galactic globular clusters, although 
the AnalySED fits and the observed colors argue for ages younger than a few
Gyr.  Both objects are marginally resolved, and would have absolute magnitudes 
of $M_{606} \approx -7.9$ and $-7.6$ at the distance of NGC 1311.  Accounting 
for the color term between F606W and the standard Johnson $V$-band, this is 
well within the range of globular cluster absolute magnitudes 
(\citeauthor{anz}\citeyear{anz}).  These are thus candidate intermediate-age 
globular star clusters.  They could possibly be foreground stars, but the 
absolute magnitudes implied by their colors ($M_{606} \approx +6$ to $+$7) 
would place them at a distance of 5 to 6 kpc.  This is highly unlikely for a 
line of sight with $b \approx -53^{\circ}$.

\subsection{Cluster Formation History}

Studies of the star formation history of low-luminosity galaxies generally 
conclude that star formation occurs in brief bursts separated by relatively 
long quiescent periods (e.g., \citeauthor{h97}\citeyear{h97}; 
\citeauthor{m98}\citeyear{m98}; \citeauthor{he04}\citeyear{he04}; 
\citeauthor{ckj}\citeyear{ckj}), during which star formation can still continue 
at a low level (\citeauthor{lvz}\citeyear{lvz}).  In the most extreme cases, 
the light from young massive star clusters dominates the bolometric output of 
ultracompact blue dwarf galaxies (e.g., \citeauthor{bcd05}\citeyear{bcd05}; 
\citeauthor{bcd06}\citeyear{bcd06}), despite the presence of much more massive
old stellar populations.  Typical late-type dwarf galaxies are also known to be 
rich in populous star clusters (\citeauthor{bhe}\citeyear{bhe}; 
\citeauthor{a04}\citeyear{a04}).  The Magellanic Clouds are our prototypes for
star-forming low-luminosity galaxies.  As such, much of our understanding of
the importance of star clusters in such galaxies follows from studies of
Magellanic Cloud clusters (e.g., \citeauthor{dGA06}\citeyear{dGA06} and
references therein).  Such studies can help us understand the factors governing 
the onset of bursts of star formation in systems that are generally not 
dominated by coherent large-scale phenomena like density waves.  Both the 
results of the detailed four-band AnalySED method 
(\citeauthor{clus03}\citeyear{clus03}; \citeauthor{clus04}\citeyear{clus04}), 
and those of the simpler SSP color-color plots argue for three epochs of 
cluster formation in NGC 1311, with ages of $\sim$10 Myr, $\sim$100 Myr, and 
$\ga$1 Gyr.  Our data for the individual stars (see Fig.~3) show the two 
younger episodes clearly, but cannot probe the oldest event, as $\ga$1 Gyr-old 
stars are too faint and red to be detected in our data.

\section{Summary and Conclusions}

NGC 1311 is a nearby, but little studied, low-luminosity late-type spiral.  It 
has optical and \hi\ properties (see Table 1) typical for such galaxies 
(\citeauthor{m98}\citeyear{m98}; \citeauthor{lvz}\citeyear{lvz}).  It is a 
member of a loose association of galaxies (\citeauthor{tully}\citeyear{tully}), 
but displays no obvious sign of any recent interaction.  We have used HST WFPC2 
and NICMOS images of NGC 1311, that span the near-UV through the near-IR, to 
identify a small population of 13 candidate star clusters.  Their masses 
increase systematically with cluster age, as would follow from the fading and 
dissolution of old, low-mass clusters, and the lack of any young super star 
clusters associated with strong starbursts.  Half the cluster candidates are 
significantly fainter than the turnover of the globular cluster luminosity 
function.  We are thus probing the range of luminosities typical of the faint 
star clusters found in Local Group dwarf galaxies, and the open cluster 
population of the Galactic disk.  Analysis of the photometry of the candidate 
star clusters suggests that NGC 1311 has had three cluster-forming episodes in 
its history, occuring $\sim$10 Myr, $\sim$100 Myr, and $\ga$1 Gyr ago.  This is 
consistent with observational work on other nearby low-luminosity galaxies 
indicating a bursting mode of star-formation.  The recent star formation, as 
traced by the NUV continuum (see Fig.~1a), is concentrated at the east and west 
ends of the central bulge-like concentration.  This is reminiscent of 
stochastic star formation models (\citeauthor{ssg}\citeyear{ssg}) as well as 
the observed properties of other low-luminosity star-forming galaxies (e.g., 
Sextans A; \citeauthor{rdp}\citeyear{rdp}).  NGC 1311 is an excellent example 
of a nearby, low-luminosity, star-forming, gas-rich galaxy that is evolving in 
relative isolation.  Understanding the star- and cluster-formation history and 
chemical evolution of such galaxies is an essential part of unraveling the 
problem of galaxy evolution.

Our next step is a study of the resolved stellar populations of NGC 1311 with 
our combined WFPC2/NICMOS HST data.  The large wavelength range of our data
allows us to sample both the very recent star formation (dominating the UV 
light) and the ancient stellar populations (from the red/near-IR light) that 
appear ubiquitous even in very late-type galaxies (e.g., 
\citeauthor{baa}\citeyear{baa}; \citeauthor{vio}\citeyear{vio}).  This should 
give us a clearer picture of the star formation history in this system, and a 
fuller understanding of the process of star formation in low-luminosity 
late-type galaxies in general.

\acknowledgments

PBE would like to thank the Department of Physics \& Astronomy at Minnesota
State University for support during this project.  We thank Andy Dolphin for
answering the first author's HSTPhot questions with patience.  We are grateful
to our referee for a report that led up to significantly improve this paper.  
This research has made use of NASA's Astrophysics Data System Bibliographic 
Services, and the NASA/IPAC Extragalactic Database (NED) which is operated by 
the Jet Propulsion Laboratory, California Institute of Technology, under 
contract with the National Aeronautics and Space Administration.  PA 
acknowledges support by the DFG grant Fr 911/11-3.  This work was supported in 
part by NASA Hubble Space Telescope grants HST-GO-09124* and HST-GO-09824* 
awarded by the Space Telescope Science Institute, which is operated by AURA for
 NASA under contract NAS 5-26555.

Facilities: \facility{HST}

\vfill
\eject

{
\begin{deluxetable}{lcc}
\tablewidth{0pt}
\tablecolumns{3}
\tablecaption{Basic Properties}
\tablehead{
\colhead{} & \colhead{} & \colhead{References}}
\startdata
$m_B$ & 13.22$\pm$0.21 & 1 \\
(B -- V) & 0.46$\pm$0.02 & 1 \\
$V_{\odot}$ & 568$\pm$5 km/sec & 2 \\
$A_V$ & 0.07 & 3 \\
$E{(B-V)}$ & 0.021 & 3 \\
D & 5.45$\pm$0.08 Mpc & 4 \\
$M_B$ & $-$15.5 & 1,4 \\
$S_{HI}$ & 15.4 Jy\,{km/sec} & 2 \\
$M_{HI}/L_B$ & 0.46 $M_{\odot} / L_{\odot}$ & 1,2,4 \\
\enddata
\tablerefs{
1) de Vaucouleurs et al.~(1991);
2) Koribalski et al.~(2004);
3) Schlegel et al.~(1998);
4) Tully et al.~(2006)
}
\end{deluxetable}
}

{
\begin{deluxetable}{cclcc}
\tablewidth{0pt}
\tablecolumns{5}
\tablecaption{Log of Observations}
\tablehead{
\colhead{Data Set} & \colhead{PID} & \colhead{Camera/Filter} & \colhead{Date} &
\colhead{Exposure}}
\startdata
          &       &              & dd-mm-yyyy & sec \\
\hline
u6dw7101m &  9124 &  WFPC2/F300W & 27-08-2001 &  300     \\
u6dw7102m &  9124 &  WFPC2/F300W & 27-08-2001 &  300     \\
u6g22403m &  9162 &  WFPC2/F606W & 22-09-2001 &  300     \\
u6g22404m &  9162 &  WFPC2/F606W & 22-09-2001 &  300     \\
u6dw7103m &  9124 &  WFPC2/F814W & 27-08-2001 &  ~40     \\
u6dw7104m &  9124 &  WFPC2/F814W & 27-08-2001 &  ~40     \\
n8ou36010 &  9824 &  NIC3/F160W  & 28-11-2003 &  512     \\
\hline
\enddata
\end{deluxetable}
}

{
\begin{deluxetable}{rccrrrrrrrrrr}
\tabletypesize{\scriptsize}
\tablewidth{0pt}
\tablecolumns{13}
\tablecaption{Photometry of Cluster Candidates}
\tablehead{
\colhead{ID} & \colhead{($X_{606}$,$Y_{606}$)} & \colhead{chip} & 
\colhead{$UV_{300}$} & \colhead{$V_{606}$} & \colhead{$I_{814}$} & 
\colhead{$H_{160}$} & \colhead{(UV$-$V)} & \colhead{(UV$-$I)} & 
\colhead{(UV$-$H)} & \colhead{(V$-$I)} & \colhead{(V$-$H)} & \colhead{(I$-$H)}}
\startdata
 & RA,Dec (J2000.0) & & $e_{UV}$ & $e_V$ & $e_I$ & $e_H$ & 
$e_{UV-V}$ & $e_{UV-I}$ & $e_{UV-H}$ & $e_{V-I}$ & $e_{V-H}$ & $e_{I-H}$ 
\\
\hline
1 & 188.20,336.93 & WF2 & 20.44 & 20.27 & 20.02 & 18.89 & 0.18 & 0.43 & 1.55 & 0.25 & 1.38 & 1.13 \\
 & 03:20:09.0,$-$52:10:39 & & 0.09 & 0.01 & 0.03 & 0.02 & 0.09 & 0.10 & 0.09 & 0.04 & 0.02 & 0.04 \\
2 & 57.97, 96.31 & WF2 & & 20.97 & & 16.75 & & & & & 4.22 & \\
 & 03:20:06.8,$-$52:10:56 & & & 0.01 & & 0.07 & & & & & 0.07 & \\
3 & 195.21,130.89 & WF3 & 18.88 & 20.25 & 20.65 & 20.52 & $-$1.38 & $-$1.78 & $-$1.64 & $-$0.40 & $-$0.27 & 0.13 \\
 & 03:20:08.0,$-$52:11:04 & & 0.04 & 0.02 & 0.11 & 0.06 & 0.05 & 0.11 & 0.08 & 0.11 & 0.07 & 0.12 \\
4 & 203.02,170.29 & WF3 & 21.51 & 22.12 & 21.42 & 19.63 & $-$0.61 & 0.09 & 1.88 & 0.70 & 2.49 & 1.79 \\
 & 03:20:08.2,$-$52:11:07 & & 0.12 & 0.02 & 0.12 & 0.03 & 0.12 & 0.17 & 0.13 & 0.12 & 0.04 & 0.13 \\
5 & 154.26,143.11 & WF3 & 20.38 & 20.78 & 20.72 & 20.03 & $-$0.40 & $-$0.33 & 0.35 & 0.07 & 0.75 & 0.69 \\
 & 03:20:07.6,$-$52:11:06 & & 0.05 & 0.01 & 0.06 & 0.04 & 0.05 & 0.08 & 0.07 & 0.06 & 0.05 & 0.07 \\
6 & 125.77,141.02 & WF3 & 21.74 & 21.50 & 21.26 & 20.15 & 0.24 & 0.49 & 1.59 & 0.24 & 1.35 & 1.11 \\
 & 03:20:07.3,$-$52:11:06 & & 0.12 & 0.01 & 0.08 & 0.05 & 0.12 & 0.14 & 0.13 & 0.08 & 0.05 & 0.09 \\
7 & 171.41,208.14 & WF3 & 21.24 & 20.12 & 19.42 & 18.30 & 1.11 & 1.81 & 2.94 & 0.70 & 1.82 & 1.12 \\
 & 03:20:07.9,$-$52:11:12 & & 0.11 & 0.01 & 0.04 & 0.02 & 0.11 & 0.12 & 0.11 & 0.04 & 0.02 & 0.04 \\
8 & 90.50,236.72 & WF3 & 20.53 & 21.42 & & & $-$0.89 & & & & & \\
 & 03:20:07.2,$-$52:11:17 & & 0.21 & 0.01 & & & 0.21 & & & & & \\
9 & 113.51,265.14 & WF3 & & 21.55 & & 19.97 & & & & & 1.58 & \\
 & 03:20:07.4,$-$52:11:19 & & & 0.01 & & 0.29 & & & & & 0.29 & \\
10 & 87.64,268.90 & WF3 & 21.94 & 20.69 & 20.14 & 18.79 & 1.25 & 1.81 & 3.15 & 0.55 & 1.90 & 1.35 \\
 & 03:20:07.2,$-$52:11:19 & & 0.16 & 0.02 & 0.08 & 0.02 & 0.16 & 0.18 & 0.16 & 0.08 & 0.03 & 0.08 \\
11 & 76.61,273.71 & WF3 & & 23.05 & & 19.43 & & & & & 3.62 & \\
 & 03:20:07.1,$-$52:11:20 & & & 0.01 & & 0.03 & & & & & 0.03 & \\
12 & 209.90, 59.05 & WF4 & 21.17 & 22.02 & 21.08 & 19.34 & $-$0.84 & 0.10 & 1.83 & 0.94 & 2.68 & 1.74 \\
 & 03:20:06.3,$-$52:11:15 & & 0.10 & 0.03 & 0.10 & 0.03 & 0.10 & 0.14 & 0.10 & 0.10 & 0.04 & 0.10 \\
13 & 200.42,111.80 & WF4 & & 21.93 & & 19.89 & & & & & 2.04 & \\
 & 03:20:05.7,$-$52:11:16 & & & 0.01 & & 0.25 & & & & & 0.25 & \\
\enddata
\end{deluxetable}
}

{
\begin{deluxetable}{rcccc}
\tablewidth{0pt}
\tablecolumns{5}
\tablecaption{Candidate Cluster Properties from AnalySED}
\tablehead{
\colhead{ID} & \colhead{$Z$} & \colhead{Age} & \colhead{$E(B-V)$} & 
\colhead{Mass}}
\startdata
 & & Myr & & 10$^3$ $M_{\odot}$ \\
\hline
 1 & 0.004 & 92$_{-84}^{+10}$ & 0.13$_{-0.05}^{+0.25}$ & 40.9$_{-29.5}^{+3.2}$ \\ 
   & 0.008 & 104$_{-67}^{+6}$ & 0.05$_{-0}^{+0.18}$ & 35.9$_{-15.9}^{+9.6}$ \\
 3 & 0.004 & 4$\pm$0 & 0.00$\pm$0 & 2.38$\pm$0 \\
   & 0.008 & 4$\pm$0 & 0.05$\pm$0 & 3.52$\pm$0.06 \\
 4 & 0.004 & 128$\pm4$ & 0.00$\pm$0 & 8.18$_{-0.05}^{+0.14}$ \\
   & 0.008 & 130$_{-6}^{+70}$ & 0.15$\pm$0.03 & 7.42$_{-0.14}^{+1.88}$ \\
 5 & 0.004 & 8$\pm$0 & 0.15$\pm$0 & 3.82$\pm$0.20 \\
   & 0.008 & 4$_{-0}^{+2}$ & 0.38$_{-0}^{+0.03}$ & 6.29$_{-0}^{+5.44}$ \\
 6 & 0.004 & 54$\pm$46 & 0.25$_{-0.15}^{+0.13}$ & 8.24$_{-4.65}^{+5.81}$ \\
   & 0.008 & 100$_{-61}^{+10}$ & 0.10$_{-0.05}^{+0.18}$ & 13.2$_{-6.90}^{+5.45}$ \\
 7 & 0.004 & 844$_{-90}^{+124}$ & 0.00$\pm$0 & 129$_{-9}^{+12}$ \\
   & 0.008 & 4$_{-0}^{+846}$ & 0.88$_{-0.88}^{+0.07}$ & 50.4$_{-4.6}^{+23.3}$ \\
10 & 0.004 & 976$_{-610}^{+150}$ & 0.00$_{-0}^{+0.40}$ & 79.7$_{-34.1}^{+2.7}$ \\
   & 0.008 & 209$_{-35}^{+1800}$ & 0.64$_{-0.64}^{+0.01}$ & 47.8$_{-14.1}^{+82.7}$ \\
12 & 0.004 & 128$_{-4}^{+0}$ & 0.00$\pm$0 & 10.5$_{-0.1}^{+0}$ \\
   & 0.008 & 12$_{-0}^{+74}$ & 0.33$_{-0.33}^{+0.02}$ & 6.43$_{-0.26}^{+4.59}$ \\
\enddata
\end{deluxetable}
}

\figcaption{42$'' \times$26$''$ images of NGC 1311 in a) F300W, b) F606W, c)
F814W, and d) F160W.  The arrow in the upper-left corner of the F300W image 
points North, the attached line-segment points East.  The scale bar in the 
lower-right corner of the F300W image is 10$''$, and applies to all four 
panels.}

\figcaption{Stellar color-color diagrams in a) $(UV_{300} - I_{814})$ 
vs.~$(V_{606} - I_{814})$, and b) $(UV_{300} - V_{606})$ vs.~$(UV_{300} - 
I_{814})$.  Solid circles with error-bars show the data.  The arrows in each 
plot show the foreground dereddening vectors.  The isochrones (Girardi et 
al.~2002) have Z=0.004, and cover a range in age from 4 Myr to 6 Gyr.}

\figcaption{F606W vs.~(F606W -- F814W) Color-Magnitude diagram with symbols and 
isochrones as in Fig.~2.  The solid lines are the 10 Myr and 100 Myr 
isochrones.  The dotted lines are younger, intervening, and older isochrones, 
sampled every 0.25 dex in age.  The points surrounded by large circles indicate 
objects with peculiar colors, as discussed in the text.  The dashed line shows 
$I_{814} = 22$.  The points surrounded by large diamonds are objects with 
peculiar colors, but fainter than this limit.  The bold arrow shows the 
foreground dereddening vector.}

\figcaption{A 48$'' \times$80$''$ region of the F606W image in which the 
photometrically selected candidate clusters are circled.}

\figcaption{Sharpness vs.~$V_{606}$.  Circled points are the photometrically 
selected candidate clusters.  Plus-signs are single-pixel objects that HSTPhot 
classifies as either hot pixels or cosmic rays.  Points enclosed in diamonds 
are classified as ``extended sources'' by HSTPhot.  The diamonds with large, 
dark crosses overlayed appear to be background galaxies based on Figure 6.}

\figcaption{The full WFPC2 F606W image, showing the photometrically selected 
candidate clusters as in Fig.~4, and additional faint cluster candidates 
enclosed in diamonds.  Note that two of the four background objects are obvious 
background spirals.}

\figcaption{$6{''}\llap{.}5$ square ``postage stamps'' of the candidate 
clusters shown in Fig.~6, not including the clear background spirals.}

\figcaption{Radial profiles of cluster candidates along with stars from the
corresponding WF chips.  The stellar profiles are the solid, bold lines. a) 
WF2 clusters 1--2, b) WF3 clusters 3--6, c) WF3 cluster 7--11 d) WF4 clusters
12-13.}

\figcaption{Histogram of observed half-light radii of candidate clusters (solid
line) along with those of the comparison stars (dotted line).}

\figcaption{Age vs.~mass from AnalySED with $Z=0.004$ for the candidate
clusters.  The line is a simple linear bisector.}

\figcaption{Color-color diagrams of the candidate clusters (large black 
squares), along with the Girardi et al.~(2002) SSP models for a range of 
metallicities (color-coded in the figure) and ages (symbols coded in the 
figure) a) $(UV_{300} - V_{606})$ vs.~$(I_{814} - H_{160})$, b) $(UV_{300} - 
V_{606})$ vs.~$(UV_{300} - H_{160})$.}


\begin{thebibliography}{}

\bibitem[Anders et al.~(2007)]{a07}
  Anders, P., Bissantz, N., Boysen, L., de Grijs, R., \& Fritze-v.~Alvensleben,
  U., 2007, \mnras, 377, 91

\bibitem[Anders et al.~(2004b)]{clus04} 
  Anders P., Bissantz N., Fritze--von Alvensleben U., \& de Grijs R., 2004b, 
  \mnras, 347, 196

\bibitem[Anders et al.~(2004a)]{a04}
  Anders, P., de Grijs, R., Fritze-v.~Alvensleben, U., \& Bissantz, N., 2004a, 
  \mnras, 347, 17

\bibitem[Anders et al.~(2006)]{apcor}
  Anders, P., Gieles, M., \& de Grijs, R., 2006, \aap, 451, 375

\bibitem[Ashman \& Zepf (1998)]{anz}
  Ashman, K.~M. \& Zepf, S.~E. 1998, Globular Cluster Systems (Cambridge, 
  Cambridge Univ.~Press).

\bibitem[Baade (1958)]{baa}
  Baade, W. 1958, in Ricerche Astronomiche, v.~5, Proceedings of a Conference 
  at Vatican Observatory, ed.~D.J.K. O'Connell (Amsterdam:  Interscience), 303

\bibitem[Bastian et al.~(2005)]{bas05}
  Bastian, N., Gieles, M., Lamers, H.~J.~G.~L.~M., Scheepmaker, R.~A., \& 
  de Grijs, R. 2005, \aap, 431, 905

\bibitem[Billett et al.~(2002)]{bhe}
  Billett, O.~H., Hunter, D.~A., \& Elmegreen, B.~G., 2002, \aj, 123, 1454

\bibitem[Binney \& Merrifield (1998)]{binmer}
  Binney, J., \& Merrifield, M., 1998, Galactic Astronomy (Princeton 
  Univ.~Press:  Princeton)

\bibitem[Calzetti (1997)]{calz}
  Calzetti, D., 1997, \aj, 113, 162

\bibitem[Cardelli et al.~(1989)]{ccm}
  Cardelli, J.~A., Clayton, G.~C., \& Mathis, J.~S. 1989, \apj, 345, 245

\bibitem[Corbin et al.~(2007)]{ckj}
  Corbin, M.~R., Kim, H., Jansen, R.~A., Windhorst, R.~A. \& Cid~Fernandes, R. 
  2007, \apj, In press

\bibitem[Corbin et al.~(2006)]{bcd06}
  Corbin, M.~R., Vacca, W.~D.,  Cid Fernandes, R., Hibbard, J.~E., Somerville, 
  R.~S., \& Windhorst, R.~A. 2006, \apj, 651, 861

\bibitem[Corbin et al.~(2005)]{bcd05}
  Corbin, M.~R., Vacca, W.~D., Hibbard, J.~E., Somerville, R.~S., \& Windhorst, 
  R.~A. 2005, \apj, 629, L89

\bibitem[de Grijs et al.~(2005)]{grijs}
  de Grijs, R., Anders, P., Lamers, H.~J.~G.~L.~M., Bastian, N., 
  Fritze-v.~Alvensleben, U., Parmentier, G., Sharina, M.~E., \& Yi, S. 2005, 
  \mnras, 359, 874

\bibitem[de Grijs \& Anders (2006)]{dGA06}
  de Grijs R. \& Anders, P. 2006, \mnras, 366, 295.

\bibitem[de Grijs et al.~(2003a)]{clus03} 
  de Grijs R., Fritze--von Alvensleben U., Anders P., Gallagher J.~S. {\sc 
  iii}, Bastian N., Taylor V.~A., \& Windhorst R.~A., 2003, \mnras, 342, 259

\bibitem[de Grijs et al.~(2003b)]{dGNA}
  de Grijs, R., Lee, J.~T., Mora Herrera, M.~C., Fritze-v.~Alvensleben, 
  U., \& Anders, P., 2003, \na, 8, 155

\bibitem[de Jong (2006)]{nonlin}
  de Jong, R.~S. 2006, ``Correcting NICMOS Count-Rate Dependent 
  Non-linearity,'' NICMOS Instrument Science Report 2006-003 (Baltimore:
  STScI).

\bibitem[de Vaucouleurs et al.~(1991)]{rc3} 
  de Vaucouleurs, G., de Vaucouleurs A., Corwin, H.~G., Jr., Buta, R.~J., 
  Paturel, G. \& Fouqu\'e, P. 1991, Third Reference Catalogue of Bright 
  Galaxies (Springer-Verlag:  New York) (RC3)

\bibitem[Dohm-Palmer et al.~(2002)]{rdp}
  Dohm-Palmer, R.~C., Skillman, E.~D., Mateo, M., Saha, A., Dolphin, A., 
  Tolstoy, E., Gallagher, J.~S., \& Cole, A.~A. 2002, \aj, 123, 813

\bibitem[Dolphin (2000)]{hstpht}
  Dolphin, A.~E. 2000, \pasp, 112, 1383

\bibitem[Dolphin \& Kennicutt (2002a)]{dk1}
  Dolphin, A.~E. \& Kennicutt, R.~C., Jr. 2002, \aj, 123, 207

\bibitem[Dolphin \& Kennicutt (2002b)]{dk2}
  Dolphin, A.~E. \& Kennicutt, R.~C., Jr. 2002, \aj, 124, 158

\bibitem[Elson et al.~(1987)]{eff87}
  Elson, R.~A.~W., Fall, S.~M., \& Freeman, K.~C. 1987, \apj, 323, 54

\bibitem[Gallart et al.~(2003)]{carme}
  Gallart, C., Zoccali, M., Bertelli, G., Chiosi, C., Demarque, P., Girardi, 
  L., Nasi, E., Woo, J.-H., Yi, S. 2003, \aj, 125, 742

\bibitem[Gieles et al.~(2005)]{gbl}
  Gieles, M., Bastian, N., Lamers, H.~J.~G.~L.~M., \& Mout, J.~N. 2005, \aap, 
  441, 949

\bibitem[Girardi et al.~(2002)]{thiso}
  Girardi, L., Bertelli, G., Bressan, A., Chiosi, C., Groenewegen, M.~A.~T., 
  Marigo, P., Salasnich, B., \& Weiss, A. 2002, \aap, 391, 195

\bibitem[Harris (1996)]{hcat}
  Harris, W.~E. 1996, \aj, 112, 1487

\bibitem[Hodge (1961)]{h61}
  Hodge, P.~W. 1961, \apj, 133, 413

\bibitem[Holtzman et al.~(1995)]{holtz}
  Holtzman, J.~A., Burrows, C.~J., Casertano, S., Hester, J.~J., Trauger, J.T.,
  Watson, A.~M., \& Worthey, G. 1995, \pasp, 107, 1065

\bibitem[Hunter (1997)]{h97}
  Hunter, D.~A. 1997, \pasp, 109, 937

\bibitem[Hunter \& Elmegreen (2004)]{he04}
  Hunter, D.~A., \& Elmegreen, B.~G. 2004, \aj, 128, 2170

\bibitem[Johnson et al.~(2001)]{j01}
  Johnson, R.~A., Beaulieu, S.~F., Gilmore, G.~F., Hurley, J., Santiago, B.~X., 
  Tanvir ,N.~R., Elson ,R.~A.~W. 2001, \mnras, 324, 367

\bibitem[King (1962)]{king62}
  King, I. 1962, \aj, 67, 471

\bibitem[Koo \& Kron (1982)]{kk82}
  Koo, D.~C. \& Kron, R.~G., 1982, \aap, 105, 107

\bibitem[Koribalski et al.~(2004)]{hipass}
  Koribalski, B.~S., Staveley-Smith, L., Kilborn, V.~A., Ryder, S.~D., 
  Kraan-Korteweg, R.~C., Ryan-Weber, E.~V., Ekers, R.~D., Jerjen, H., Henning, 
  P.~A., Putman, M.~E., Zwaan, M.~A., de Blok, W.~J.~G., Calabretta, M.~R., 
  Disney, M.~J., Minchin, R.~F., Bhathal, R., Boyce, P.~J., Drinkwater, M.~J., 
  Freeman, K.~C., Gibson, B.~K., Green, A.~J., Haynes, R.~F., Juraszek, S., 
  Kesteven, M.~J., Knezek, P.~M., Mader, S., Marquarding, M., Meyer, M., Mould, 
  J.~R., Oosterloo, T., O'Brien, J., Price, R.~M., Sadler, E.~M., Schr\"oder, 
  A., Stewart, I.~M., Stootman, F., Waugh, M., Warren, B.~E., Webster, R.~L., 
  \& Wright, A.~E. 2004, \aj, 128, 16

\bibitem[Larsen (2004)]{lar04}
  Larsen, S.~S. 2004, \aap, 416, 537

\bibitem[Mackey \& Gilmore (2003)]{mg03}
  Mackey, A.~D. \& Gilmore, G.~F., 2003, \mnras, 338, 85

\bibitem[Maraston (2005)]{m05}
  Maraston, C. 2005, \mnras, 362, 799

\bibitem[Mateo (1998)]{m98}
  Mateo, M.~L. 1998, \araa, 36, 435

\bibitem[Mora et al.~(2007)]{m07}
  Mora, M.~D., Larsen, S.~S., \& Kissler-Patig, M. 2007, \aap, 464, 495

\bibitem[Noll et al.~2004]{nic70}
  Noll, K. et al.~2004 "NICMOS Instrument Handbook", Version 7.0, (Baltimore: 
  STScI). 

\bibitem[Persson et al.~(1983)]{p83}
  Persson, S.~E., Aaronson, M., Cohen, J.~G., Frogel, J.~A. \& Matthews, K. 
  1983, \apj, 266, 105

\bibitem[Ratnatunga \& Bahcall (1985)]{rbm}
  Ratnatunga, K.~U. \& Bahcall, J.~N., 1985, \apjs, 59, 63

\bibitem[Schlegel et al.~(1998)]{sfd98}
  Schlegel, D.~J., Finkbeiner, D.~P. \& Davis, M. 1998, \apj, 500, 525

\bibitem[Seiden et al.~(1979)]{ssg}
  Seiden, P.~E., Schulman, L.~S., \& Gerola, H. 1979, \apj, 232, 702

\bibitem[Seth et al.~(2004)]{seth04}
  Seth, A., Olsen, K., Miller, B., Lotz, J. \& Telford, R. 2004, \aj, 127, 798.

\bibitem[Stetson (1987)]{hat87}
  Stetson, P.~B. 1987, \pasp, 99, 191

\bibitem[Stover (1988)]{stov88}
  Stover, R.~J., 1988, in ``Instrumentation for Ground-based Optical Astronomy: 
  Present and Future,'' ed.~L.B.~Robinson (Springer:  New York), p.~443

\bibitem[Taylor et al.~(2005)]{vio}
  Taylor, V.~A., Jansen, R.~A., Windhorst, R.~A., Odewahn, S.~C. \& Hibbard, 
  J.~E. 2005, \apj, 630, 784

\bibitem[Tully et al.~(2006)]{tully}
  Tully, R.~B., Rizzi, L., Dolphin, A.~E., Karachentsev, I.~D., Karachentseva, 
  V.~E., Makarov, D.~I., Makarova, L., Sakai, S., \& Shaya, E.~J., 2006 \aj,
  132, 729

\bibitem[van Zee (2001)]{lvz}
  van Zee, L. 2001, \aj, 121, 2003

\bibitem[Whitmore et al.~(1999)]{whit99}
  Whitmore, B.~C., Qing, Z., Leitherer, C., Fall, S.~M., Schweizer, F. \& 
  Miller, B.~W. 1999, \aj, 118, 1551.

\bibitem[Windhorst et al.~(2002)]{wind02}
  Windhorst, R.~A., Taylor, V.~A., Jansen, R.~A., Odewahn S.~C., Chiarenza, 
  C.~A.~T., de Grijs, R., de Jong, R.~S., Frogel, J.~A., Eskridge, P.~B., 
  Gallagher, J.~S., Conselice, C., Hibbard, J.~E., Matthews, L.~D., MacKenty, 
  J. \& O'Connell, R.~W. 2002, \apjs, 143, 113

\end{thebibliography}
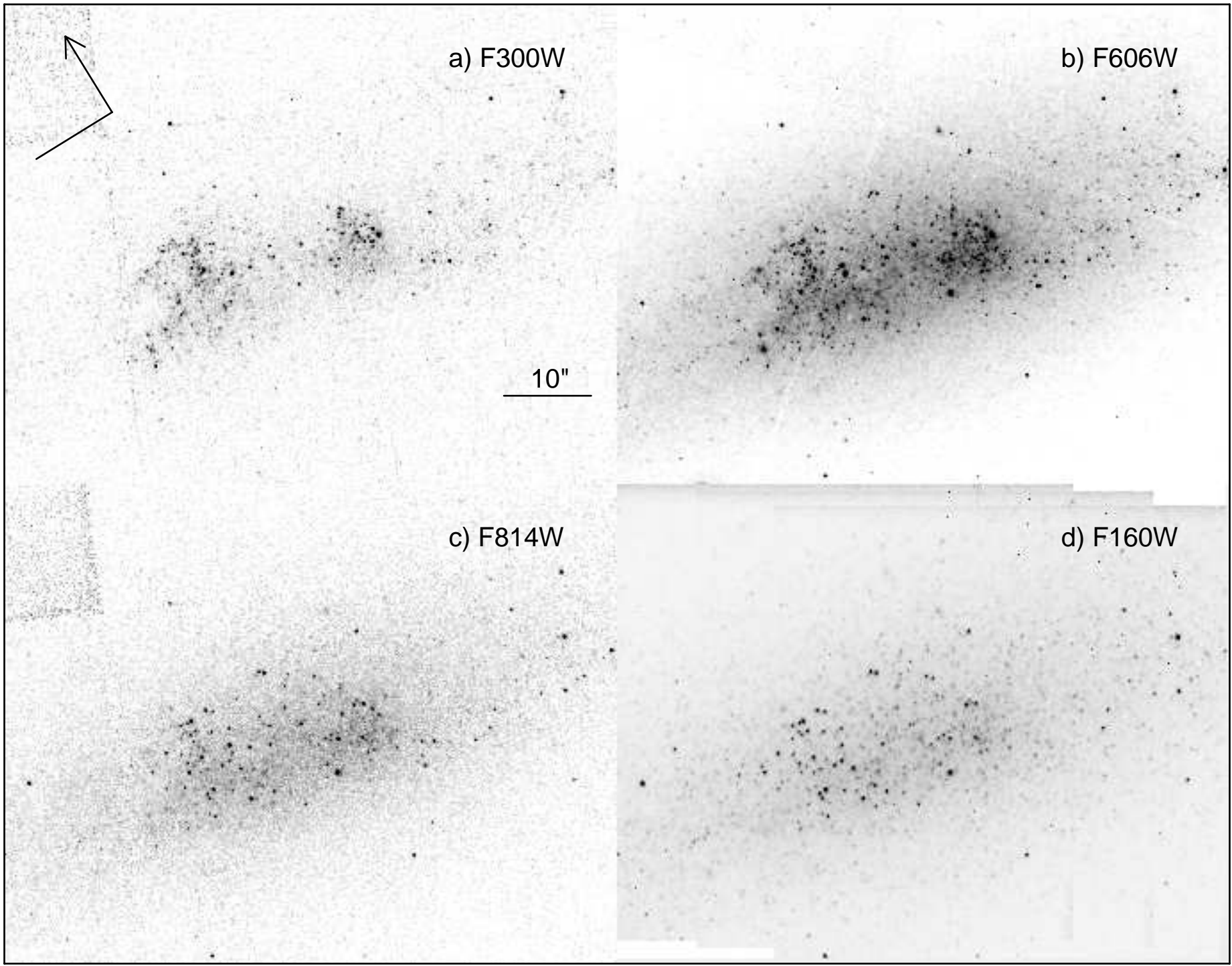
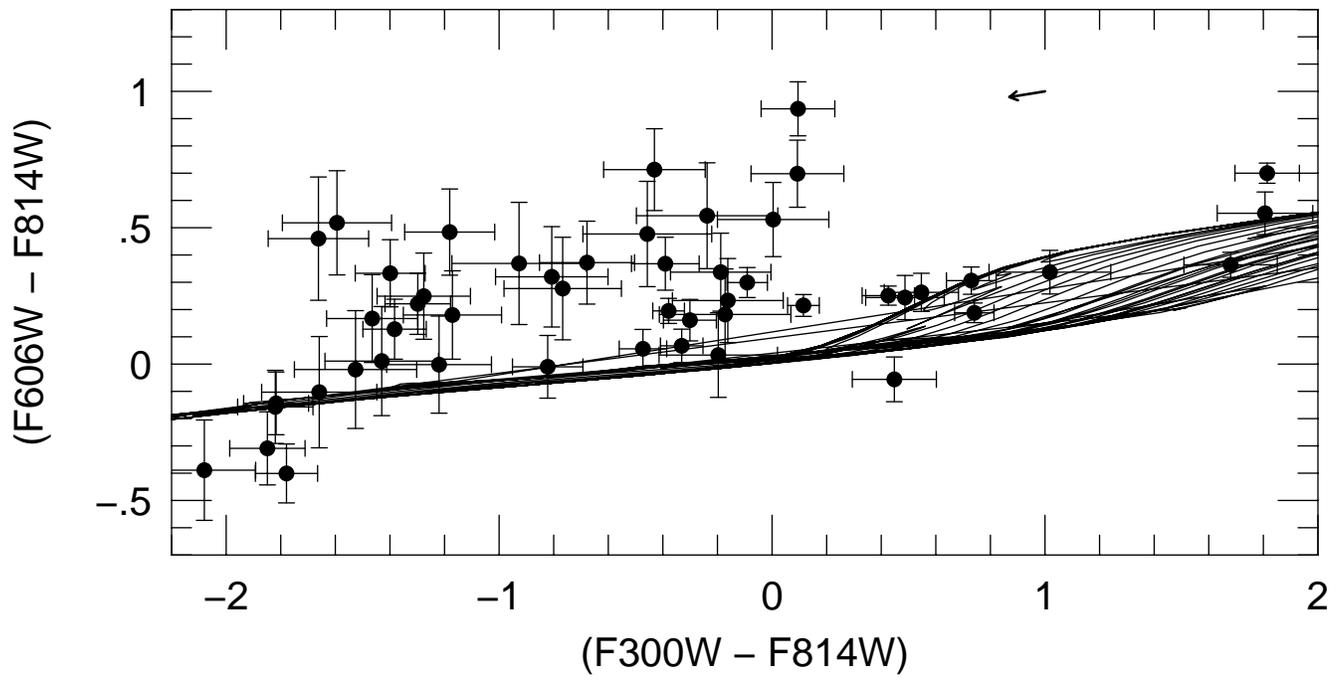
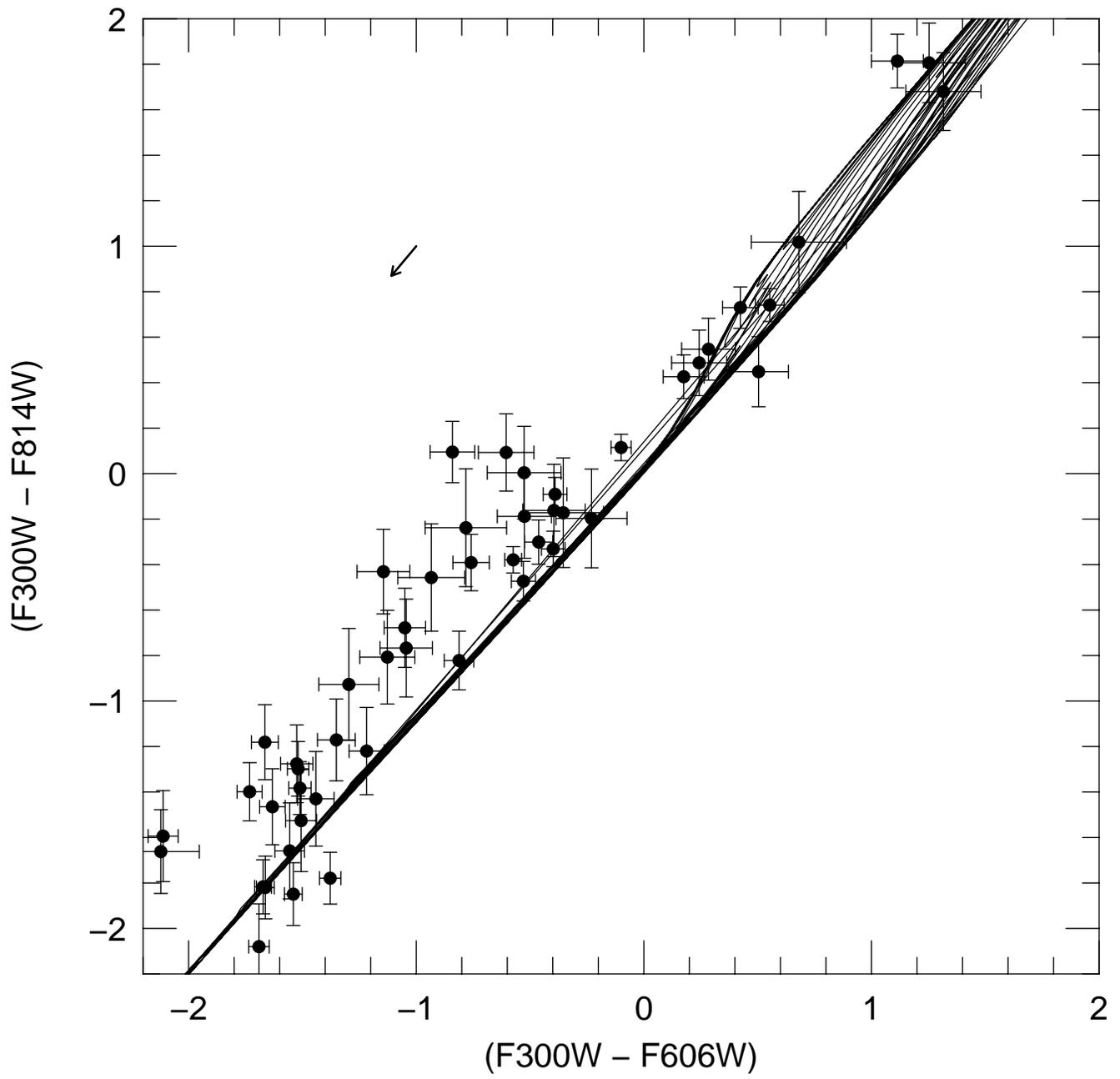
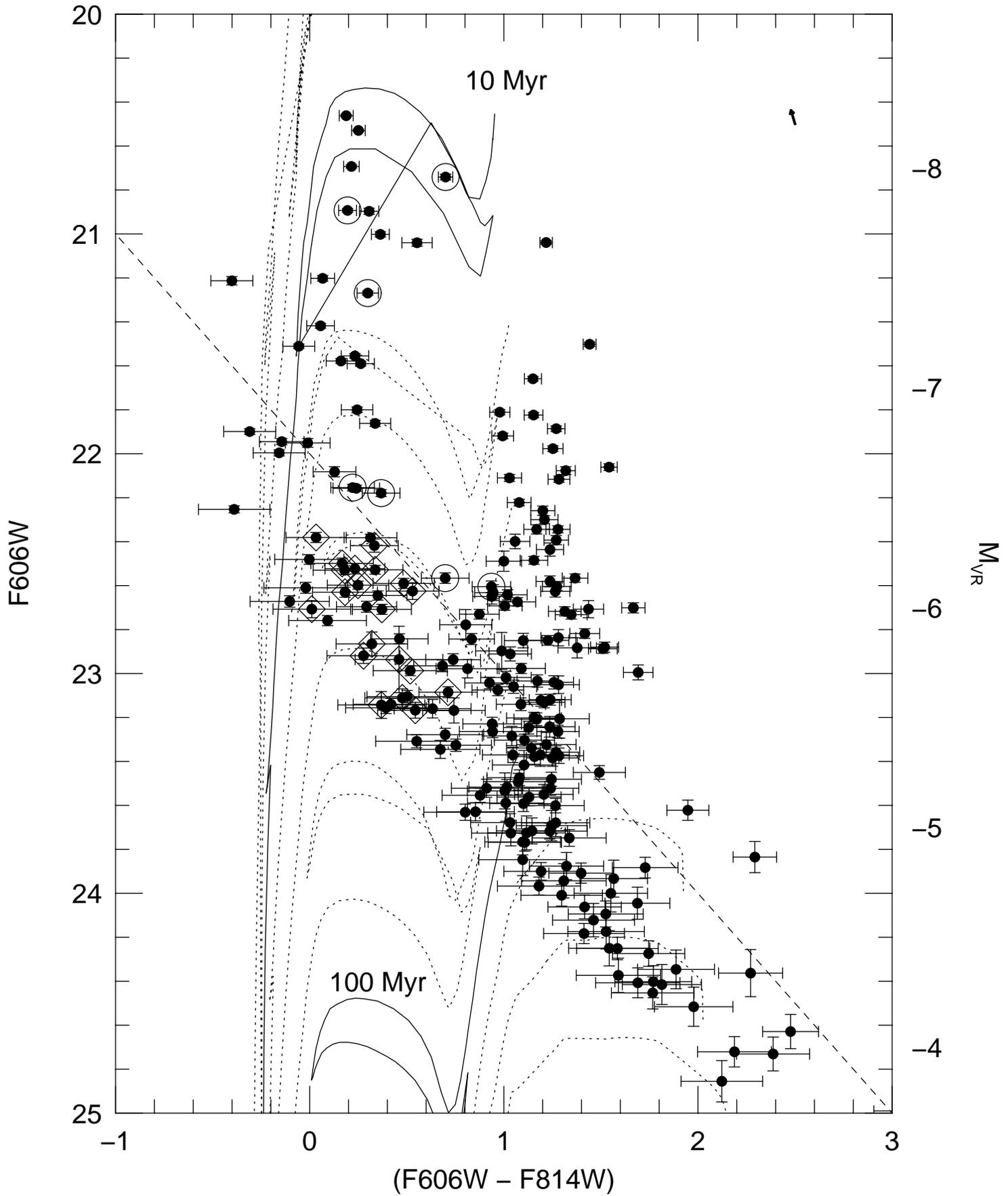
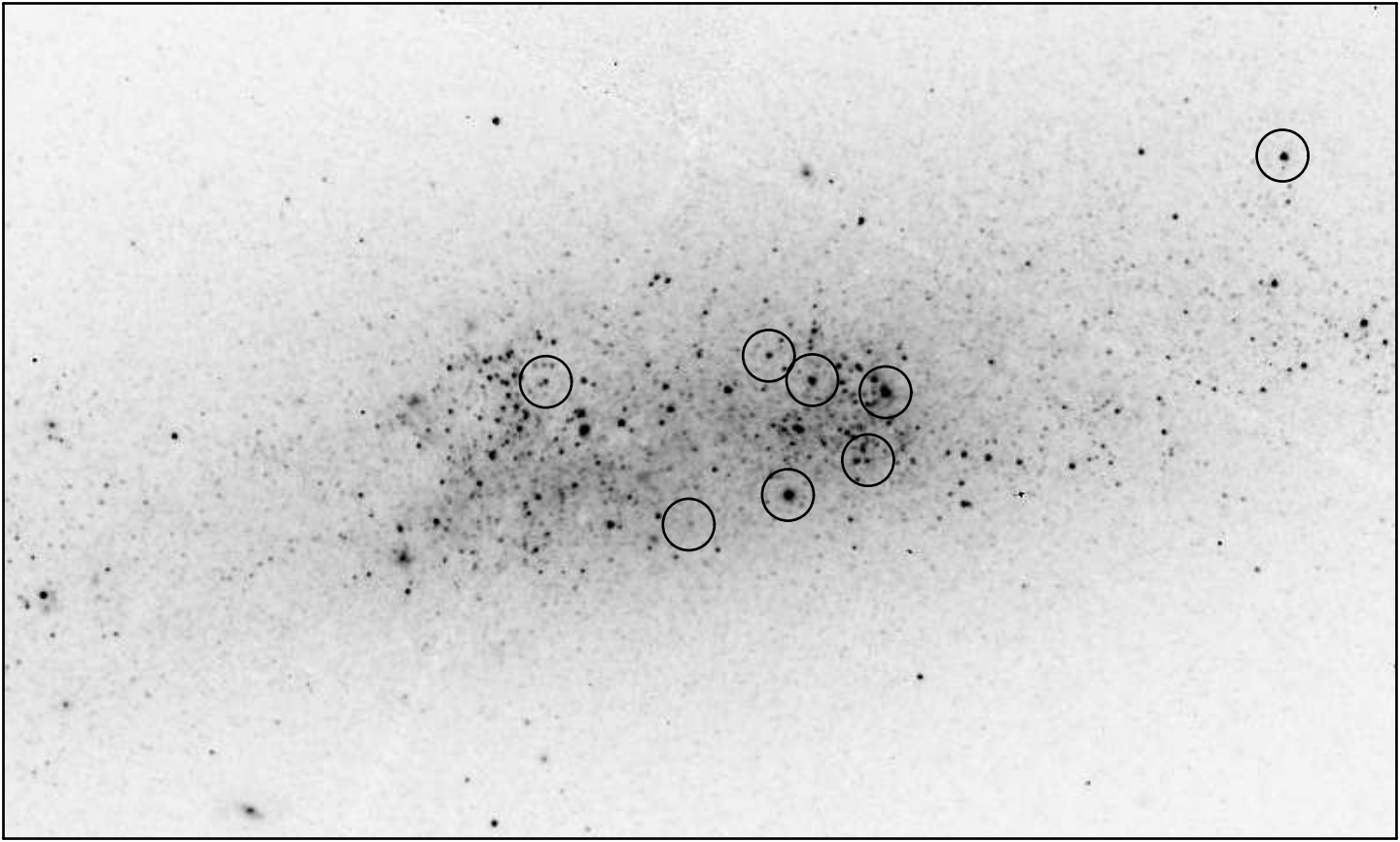
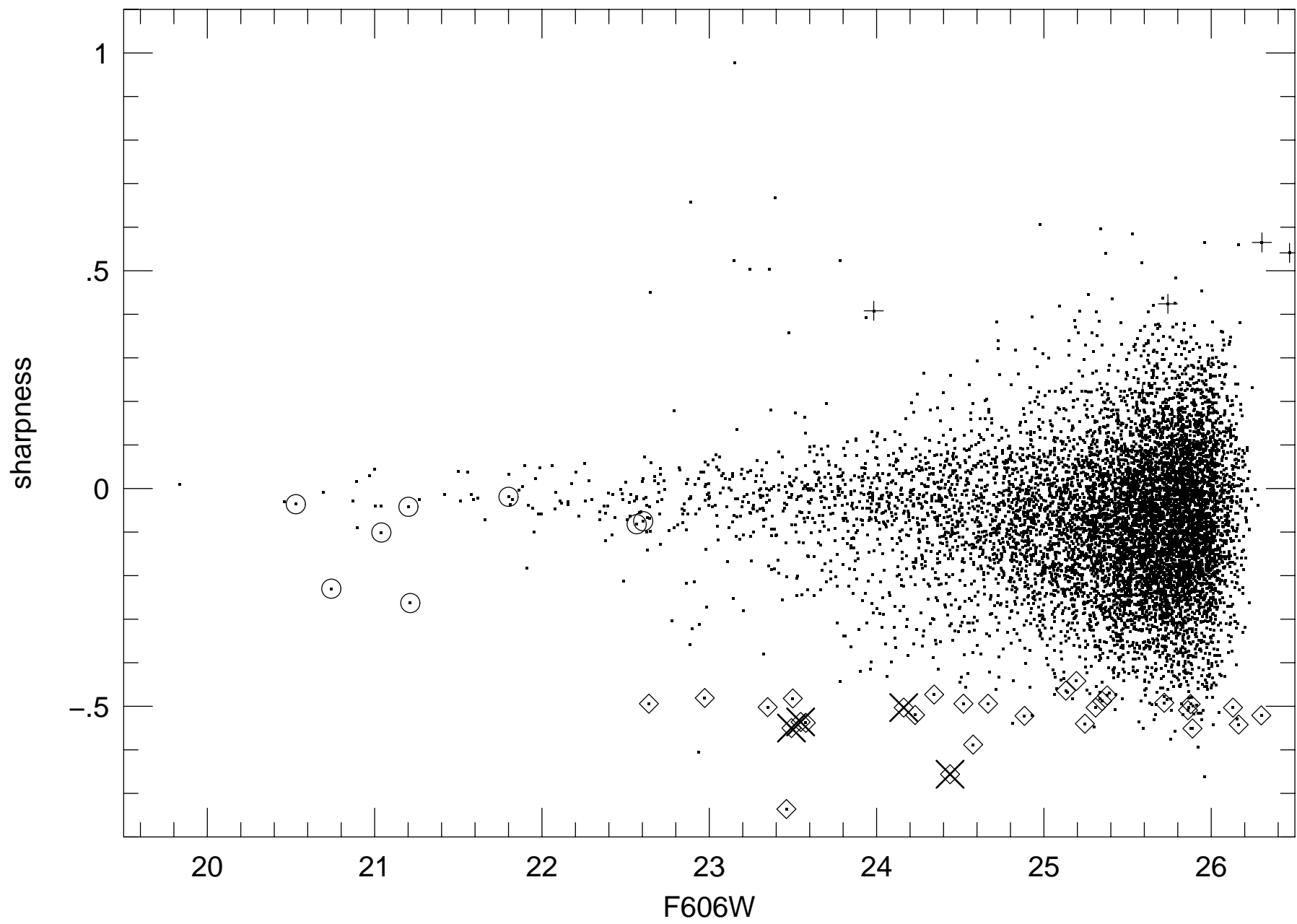
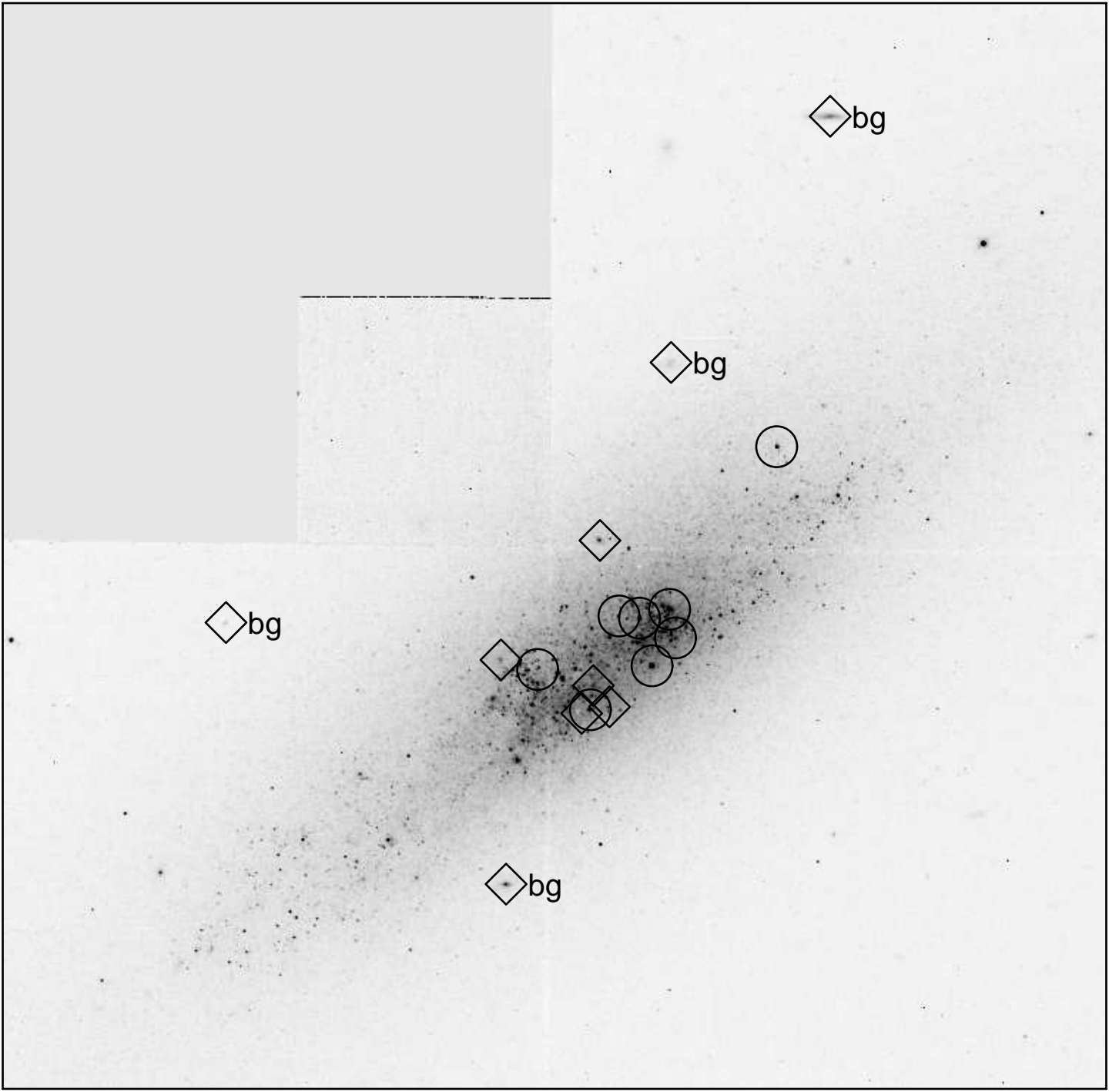
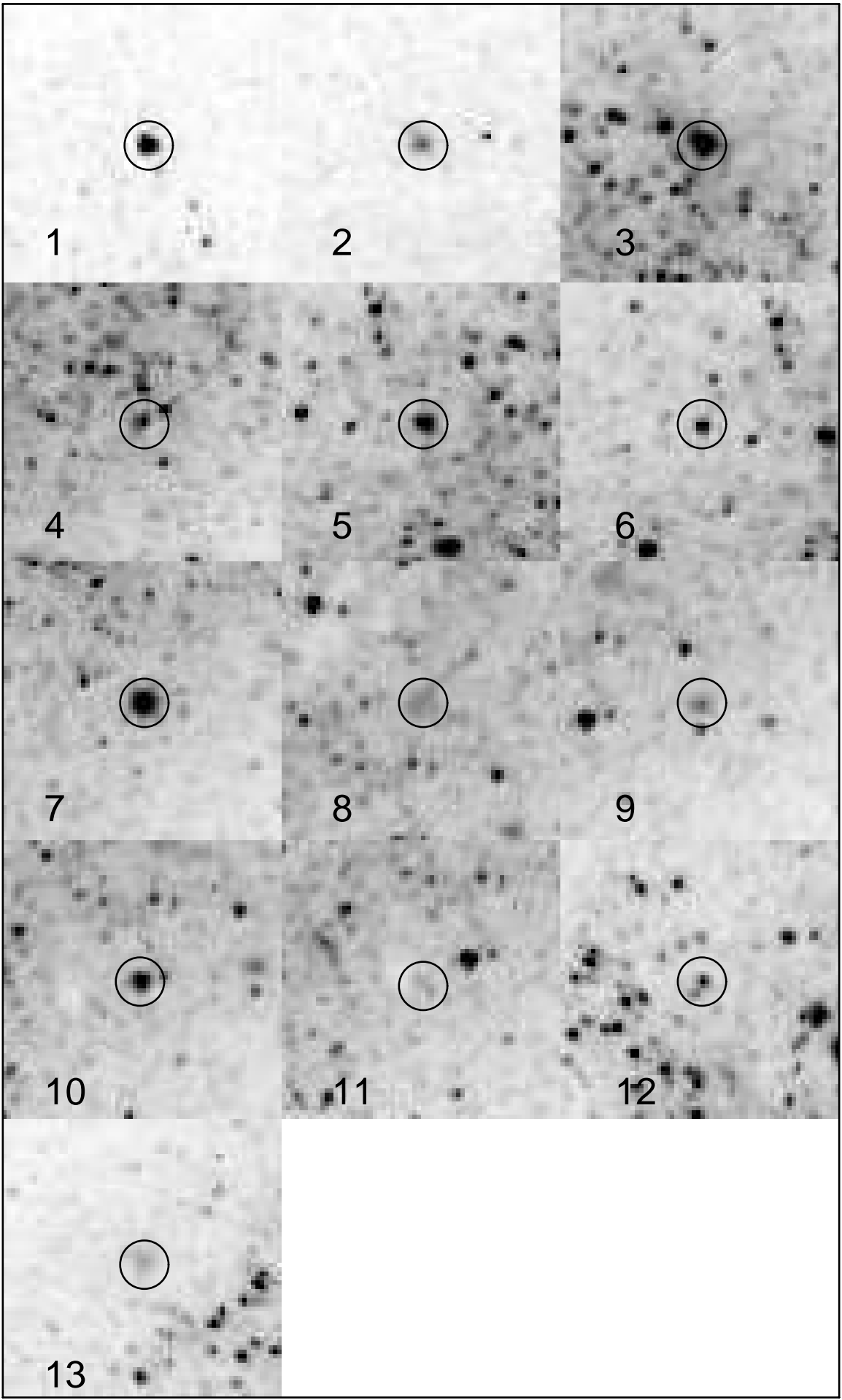
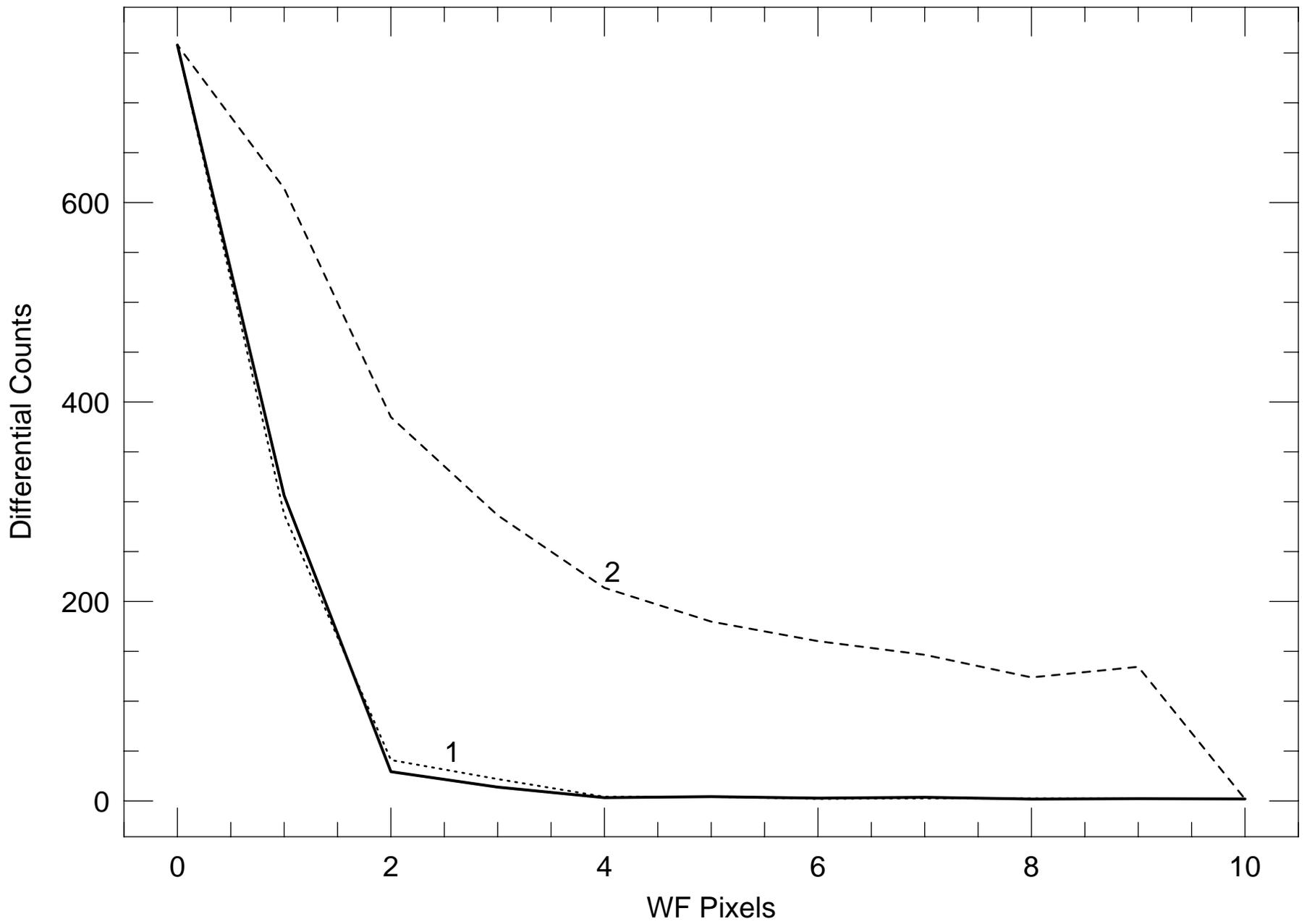
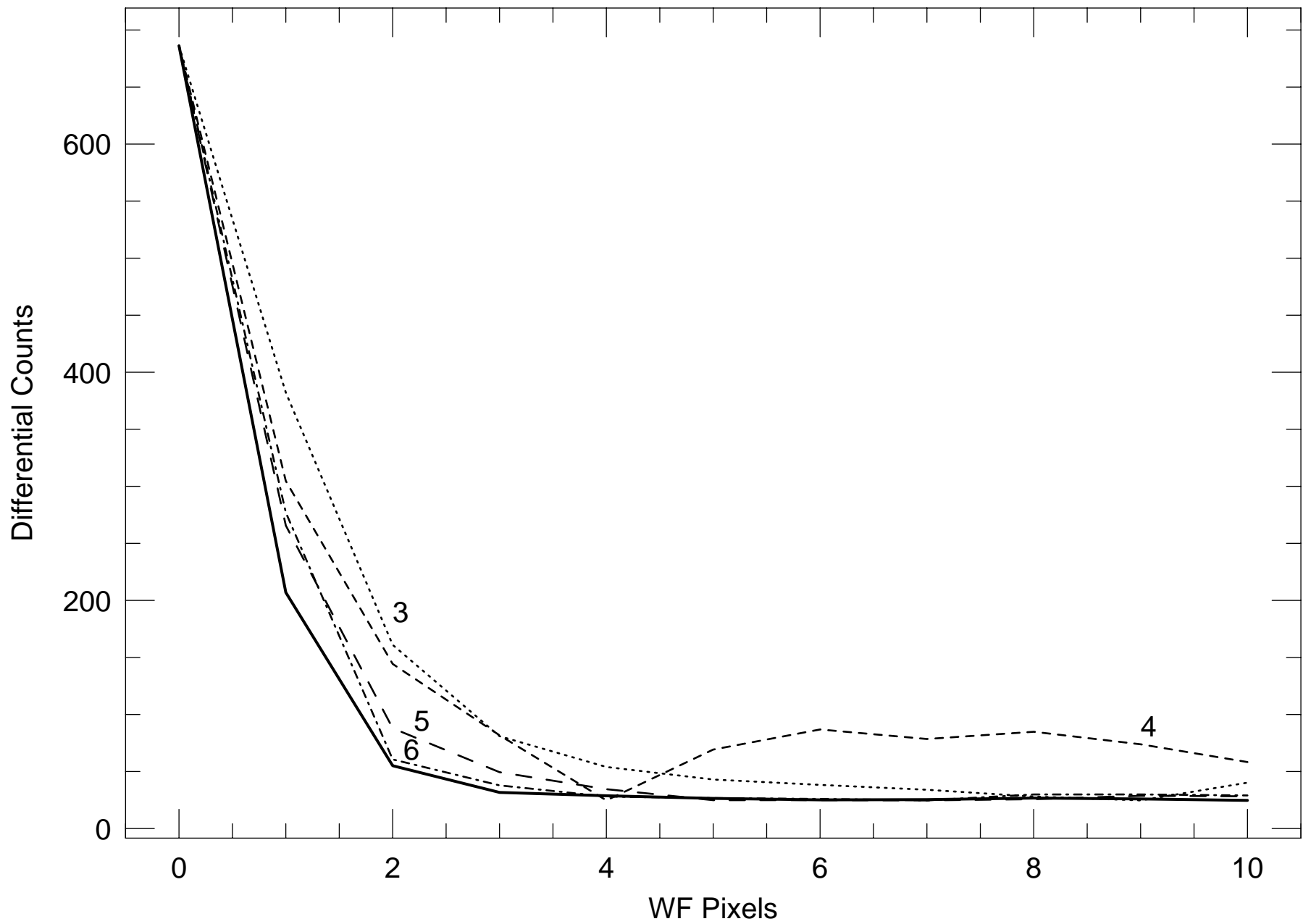
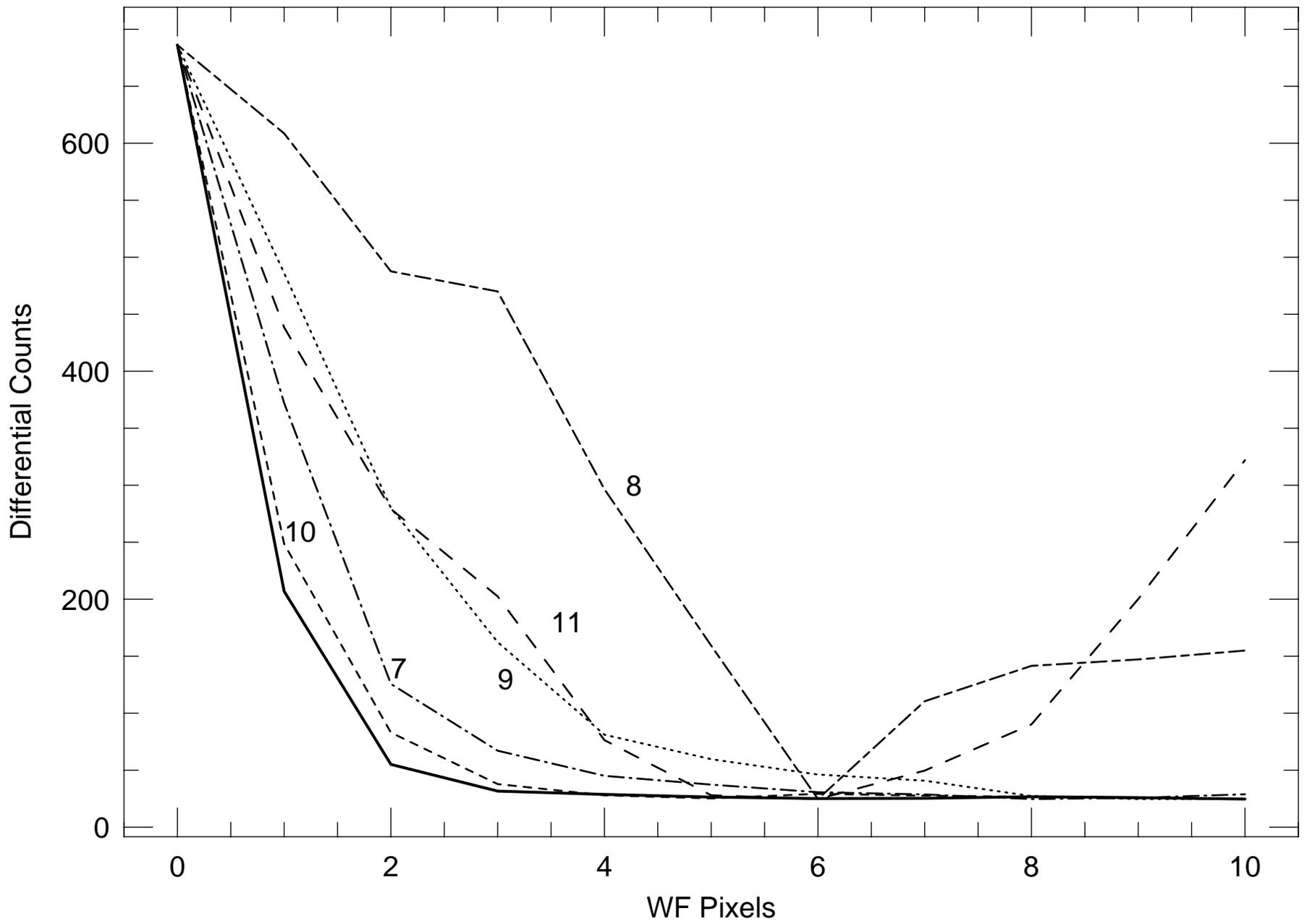
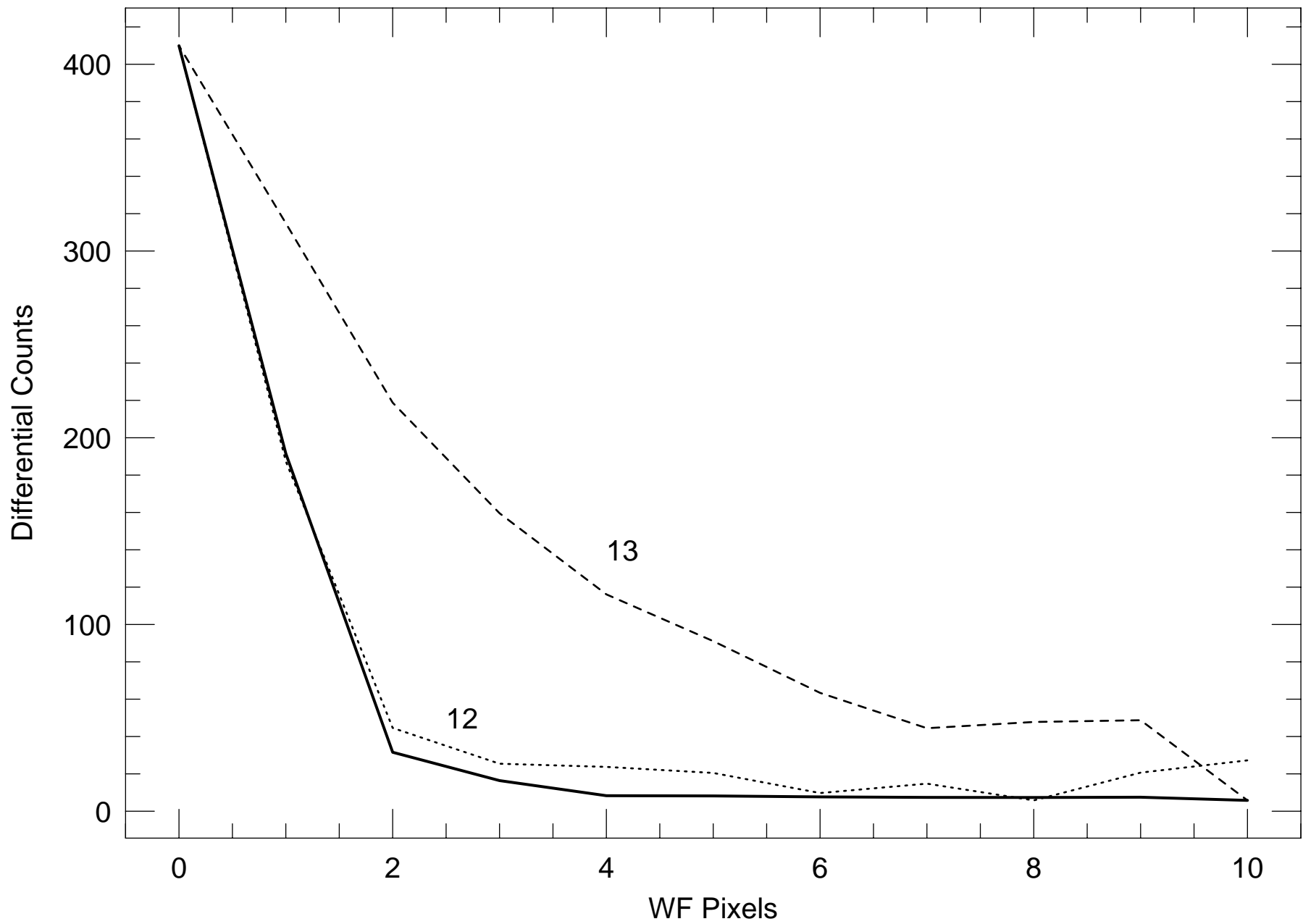
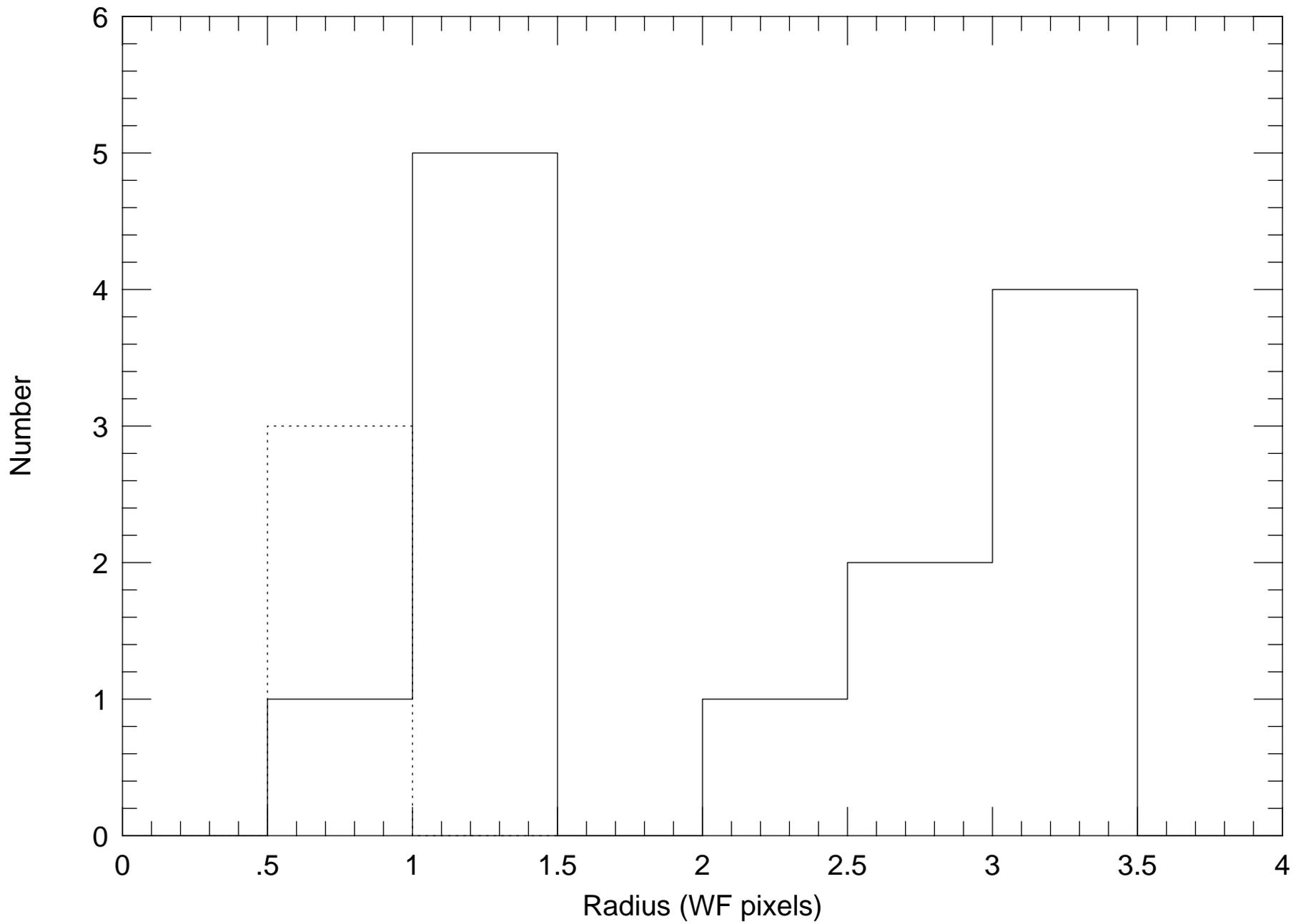
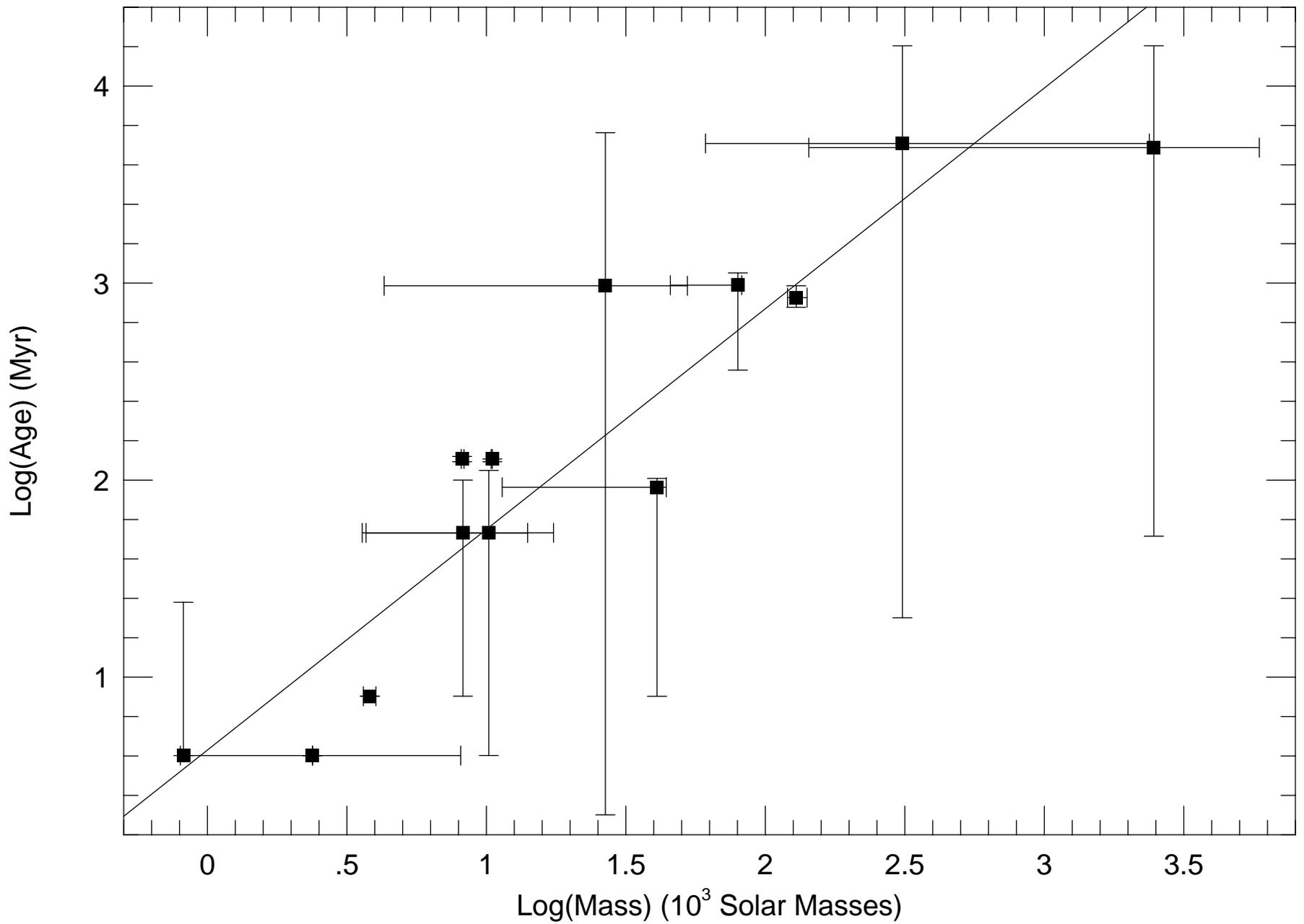
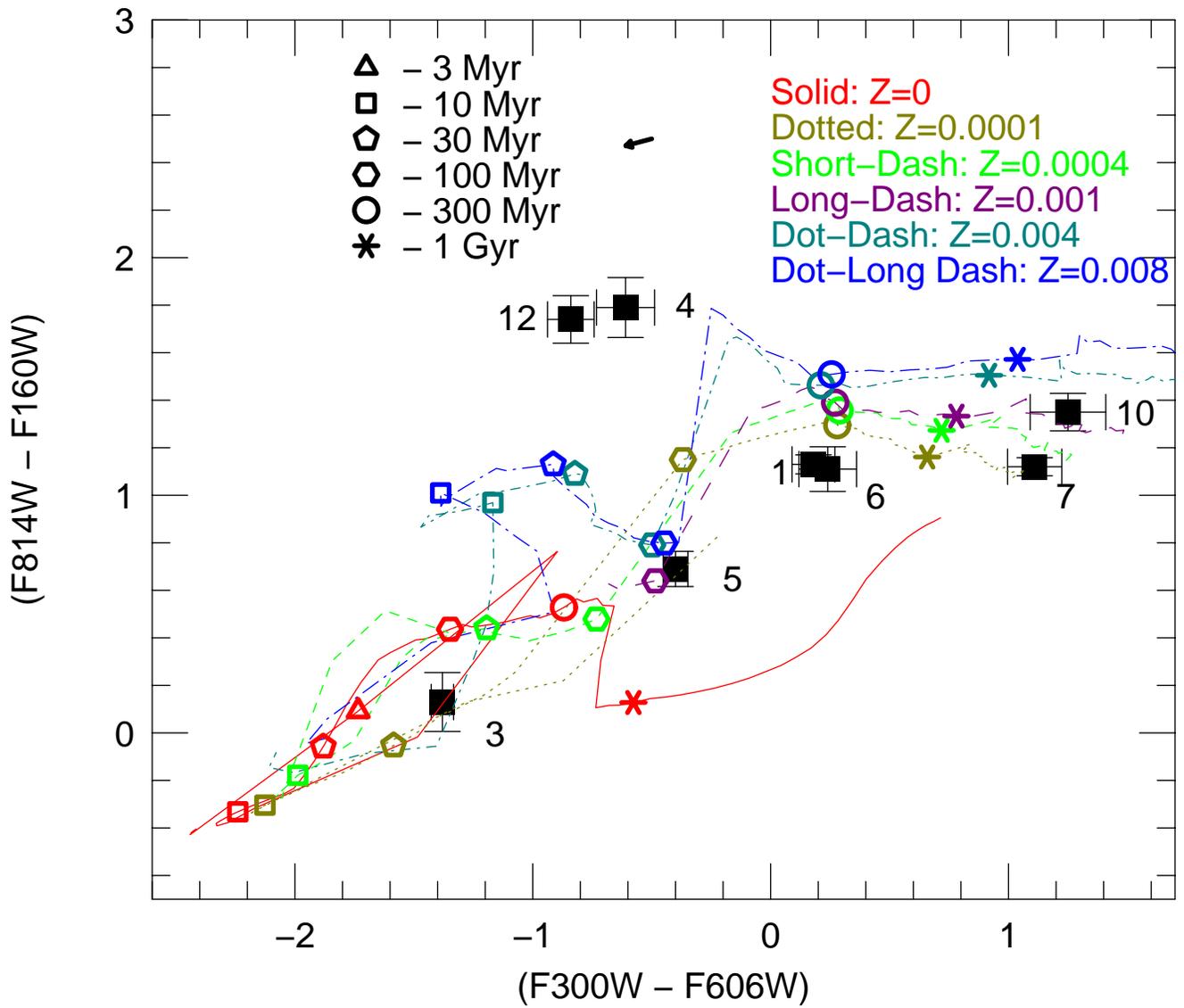
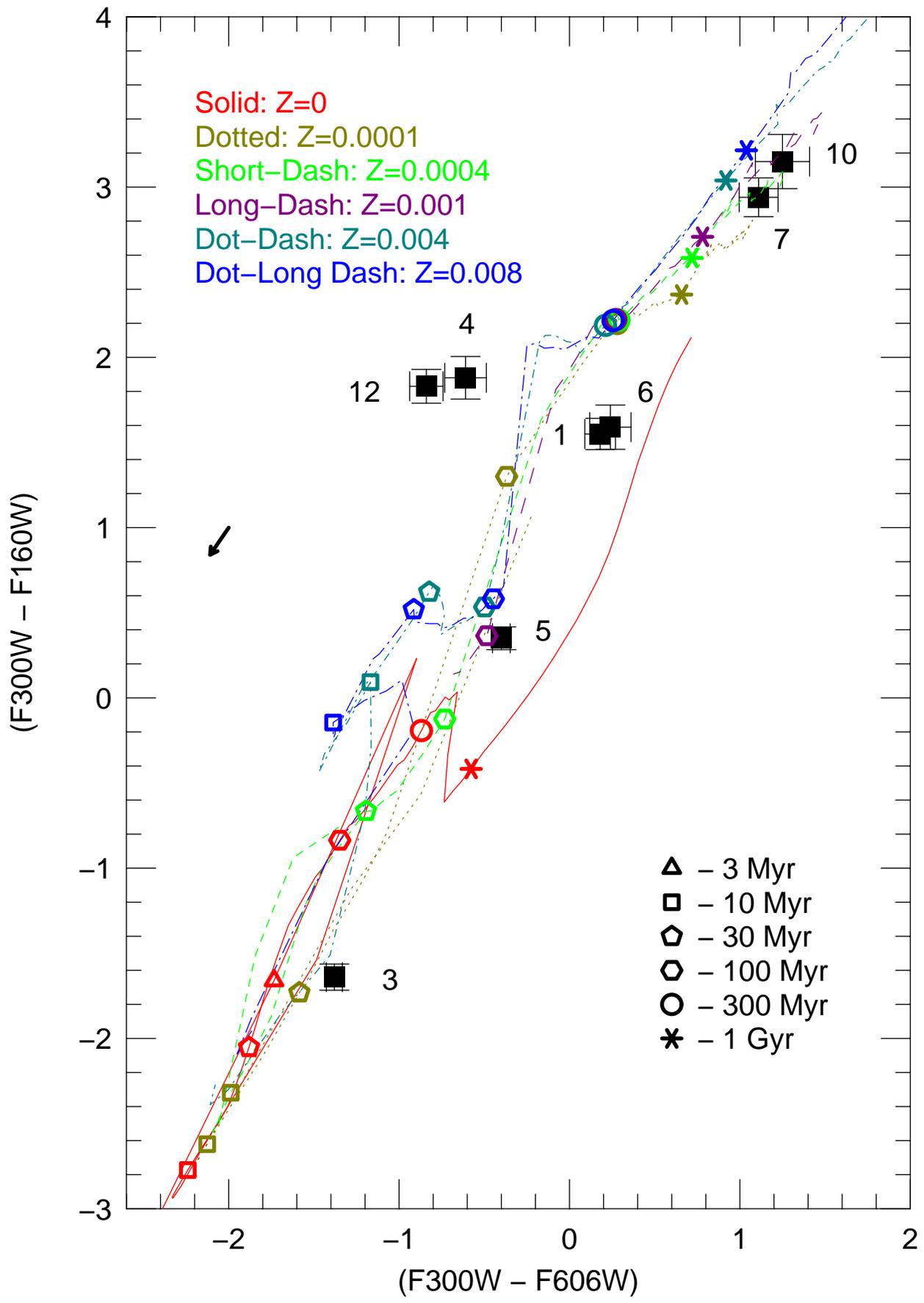
\end{document}